\documentclass[aps,letterpaper,twocolumn,preprintnumbers,floatfix,superscriptaddress]{revtex4}

\usepackage{graphicx}
\usepackage{dcolumn}
\usepackage{bm}
\usepackage{epsfig}
\usepackage{gensymb}
\usepackage{mathrsfs}
\usepackage{extpfeil}
\usepackage{extarrows}
\usepackage{float}
\usepackage{amsmath}
\usepackage{amssymb}
\usepackage{graphicx,bm}
\usepackage{natbib}
\usepackage{slashed}
\usepackage[makeroom]{cancel}
\usepackage{dsfont}
\usepackage{longtable}
\usepackage{multirow}
\usepackage{youngtab}
\usepackage{young}
\usepackage{tikz}
 \usepackage[titletoc]{appendix}

\def\fun#1#2{\lower3.6pt\vbox{\baselineskip0pt\lineskip.9pt
  \ialign{$\mathsurround=0pt#1\hfil##\hfil$\crcr#2\crcr\sim\crcr}}}

\usepackage{tikz}
\usetikzlibrary{decorations.pathmorphing,decorations.markings,trees,shapes}

\def\lsim{\mathrel{\rlap{\raise 2.5pt \hbox{$<$}}\lower 2.5pt\hbox{$\sim$}}}
\def\gsim{\mathrel{\rlap{\raise 2.5pt \hbox{$>$}}\lower 2.5pt\hbox{$\sim$}}}

\input epsf

\newcommand{\blue}[1]{\textcolor{blue}{#1}}
\usepackage{color}

\newcommand{\comment}[1]{}

\newcommand{\be}{\begin{equation}}
\newcommand{\ee}{\end{equation}}
\newcommand{\bea}{\begin{eqnarray}}
\newcommand{\eea}{\end{eqnarray}}

\newcommand{\vev}[1]{\langle #1 \rangle}
\newcommand{\ket}[1]{| #1 \rangle}

\newcommand{\antiket}[1]{| #1 ]}

\newcommand{\eq}[1]{\begin{equation}\begin{split} #1 \end{split}\end{equation}}

\renewcommand\arraystretch{2}

\begin{document}

\title{Constructing Generic Effective Field Theory for All Masses and Spins}

\author{Zi-Yu Dong}
\affiliation{CAS Key Laboratory of Theoretical Physics, Institute of Theoretical Physics,
Chinese Academy of Sciences, Beijing 100190, China.}
\affiliation{School of Physical Sciences, University of Chinese Academy of Sciences, Beijing 100190, P. R. China.}
\affiliation{Department of Physics, LEPP, Cornell University, Ithaca, NY 14853, USA}
\author{Teng Ma}
\affiliation{Physics Department, Technion -- Israel Institute of Technology, Haifa 3200003, Israel}
\author{Jing Shu}
 \affiliation{CAS Key Laboratory of Theoretical Physics, Institute of Theoretical Physics,
Chinese Academy of Sciences, Beijing 100190, China.}
\affiliation{School of Physical Sciences, University of Chinese Academy of Sciences, Beijing 100190, P. R. China.}
    \affiliation{School of Fundamental Physics and Mathematical Sciences, Hangzhou Institute for Advanced Study, University of Chinese Academy of Sciences, Hangzhou 310024, China}
    \affiliation{International Center for Theoretical Physics Asia-Pacific, Beijing/Hanzhou, China}

\author{Yu-Hui Zheng}
\affiliation{CAS Key Laboratory of Theoretical Physics, Institute of Theoretical Physics,
Chinese Academy of Sciences, Beijing 100190, China.}
\affiliation{School of Physical Sciences, University of Chinese Academy of Sciences, Beijing 100190, P. R. China.}

\begin{abstract}
We fully solve the long-standing problem of operator basis construction for fields with any masses and spins. Based on the on-shell method, we propose a novel method to systematically construct a complete set of lowest dimensional  amplitude bases at any given dimension through semi-standard Young tableaus of Lorentz subgroup $SU(2)_r$ and global symmetry $U(N)$ ($N$ is the number of external legs), which can be directly mapped into physical operator bases. We first construct a complete set of monomial bases whose dimension is not the lowest and a redundant set of bases that always contains a complete set of amplitude bases with the lowest dimension. Then we decompose the bases of the redundant set into the monomial bases from low to high dimension and eliminate the linear correlation bases. Finally, the bases with the lowest dimension can be picked up. We also propose a matrix projection method to construct the massive amplitude bases involving identical particles. The operator bases of a generic massive effective field theory can be efficiently constructed by the computer programs. A complete set of four-vector operators at dimensions up to six is presented.
\end{abstract}

\pacs{xxx}

\maketitle

\section{Introduction}
Effective field theory (EFT) of massive fields is widely applied in particle physics, such as lower energy QCD~\cite{Weinberg:1966kf,Weinberg:1968de, Weinberg:1978kz, Gasser:1982ap,Gasser:1983yg}, Higgs EFT (HEFT)~\cite{Falkowski:2019tft,Cohen:2020xca}, dark matter EFT~\cite{Goodman:2010ku,Cao:2009uw,Zheng:2010js,Aebischer:2022wnl}, and low energy EFT~\cite{Jenkins:2017jig}. Compared with massless EFTs, massive EFTs have many advantages in new physics (NP) study.  For example, HEFT can fully describe the IR effects of the NP models in which electroweak symmetry is non-linearly realized, but standard model EFT (SMEFT) can not~\cite{Falkowski:2019tft,Cohen:2020xca}. Massive EFT is more convenient for studying IR effects of NP theory at the electroweak symmetry breaking (EWSB) phase. For example, there is no field normalization issue, and generally, low point (three- and four-point) operator bases are enough for most low-energy phenomenology studies.

However, constructing a generic massive EFT is still a long-standing problem.
A complete set of EFT bases is essential to fully categorize and parametrize the infrared (IR) effects of any ultraviolet (UV) theory. Nevertheless, in traditional field theory, constructing independent EFT bases is challenging because of operator redundancy from equation of motion (EOM) and integration by part (IBP).   

On-shell scattering amplitude is efficient in dealing with some problems of EFT, such as calculating the running of EFT operators~\cite{Bern:2020ikv,Jiang:2020mhe,EliasMiro:2020tdv,Baratella:2020lzz, Baratella:2020dvw,AccettulliHuber:2021uoa,Shu:2021qlr}, deriving EFT selecting rules~\cite{Cheung:2015aba,Jiang:2020rwz,Rose:2022njd}, and constructing scalar EFT with non-trivial soft-limit~\cite{Cheung:2014dqa,Cheung:2016drk,Low:2014nga,Low:2014oga}. Especially it is very efficient in constructing EFT bases of massless fields (called amplitude bases)~\cite{Elvang:2010jv,Shadmi:2018xan,Ma:2019gtx,Falkowski:2019zdo,AccettulliHuber:2021uoa}. A complete set of the amplitude bases without IBP and EOM redundancy can be systematically constructed by the semi-standard Young tableaus (SSYTs) of the global symmetry of massless spinors~\cite{Henning:2019enq} (more applications can be found in~\cite{Li:2020gnx,Li:2020zfq}).

However, this method is not applicable in constructing amplitude bases for massive fields. In massive EFT, besides the issues of EOM  and IBP redundancy, the dimension of massive amplitude bases should be minimized so that they can describe the leading IR effects of a UV theory (some primary explorations can be found in~\cite{Durieux:2019eor,Durieux:2020gip,Falkowski:2020fsu,Balkin:2021dko}.). Therefore in this work, we propose a novel method to systematically construct a complete set of amplitude bases with the lowest dimension based on our previous work~\cite{Dong:2021yak}. We first construct a complete set of simplified amplitude bases through the SSYTs of Lorentz subgroup $SU(2)_r$ ($SO(3,1) \simeq SU(2)_l \times SU(2)_r $) and global symmetry $U(N)$ ($N$ is the number of external particles), and then systematically construct an over redundant set of amplitude bases that always contains a complete set of amplitude bases with the lowest dimension based on polarization tensor classification. Then we decompose the bases of this redundant set from low to high dimension into the simplified amplitude bases and eliminate the linear correlation bases. Finally, the complete set of bases with the lowest dimension can be picked up. We also prove that the leading order decomposition without including the terms containing mass factors is enough to determine the independence of bases. So the decomposition can be very efficient, and a complete set of operator bases at any dimension can be easily constructed.

Within the framework of this theory, we propose a matrix projection method to get the amplitude bases involving identical particles. Instead of constructing the gauge structure and kinematic part separately~\cite{Li:2020gnx}, we act the matrix representation of Young Operator on a complete set of amplitude bases, and the bases satisfying Bose/Fermi statistics can be projected out. Based on our theory, we write a Mathematica code that can construct a complete set of massive amplitude bases at any dimension and explicitly list all bases of four massive vectors at dimension-four and six.

The paper is organized as follows. Sec.~\ref{sec:BHbase} first reviews the theory proposed in~\cite{Dong:2021yak} for massive amplitude base construction. Then explain how to construct the simplified amplitude bases and prove their independence and completeness. Sec.~\ref{sec:decomposition} illustrates how to decompose a polynomial of spinor products into a complete set of simplified amplitude bases. Sec.~\ref{sec:CFbase} discusses how to systematically construct the redundant set of amplitude bases that contains a complete set of amplitude bases with minimal dimension and gives an example to explain how to obtain the minimal dimension bases. Sec.~\ref{sec:identical} briefly discusses how to construct amplitude bases involving identical particles. We conclude in Sec.~\ref{sec:conclusion}. The appendices explain how to deal with identical particles and some examples for amplitude base construction.  We also list all the independent operator bases of four vectors at dimension-four and six.

\section{Simplification of EFT Amplitude Basis}
\label{sec:BHbase}
In this section, we will first review the theory on constructing a complete set of amplitude bases proposed in~\cite{Dong:2021yak} and then explain how to obtain the simplified amplitude bases, which are just monomials of spinor products.

\subsection{Independent Amplitude Bases}
As discussed in~\cite{Dong:2021yak}, the Lorentz structure of any scattering amplitude with $m$ massive and $n$ massless legs can be factorized as
\eq{ \label{eq:AG}
\mathcal{M}^I_{m,n}=\sum_{\{\dot{\alpha}\}} \mathcal{A}_{\{\dot{\alpha}\}}
^{\{I_{1,\cdots\!\,,2s_i}^{i} \}} (\epsilon_i)G^{\{\dot{\alpha}\}}(|j],|j\rangle,p_i),
}
where massive little group tensor structure (MLGTS) $\mathcal{A}^{\{I\}}\sim \bigotimes_{i=1}^m\epsilon_i$ is the linear function of
$\epsilon_i \equiv \antiket{i}^{\{ I_{1}}_{\dot \alpha_1},\ldots, \antiket{i}^{I_{2s_i} \}}_{\dot\alpha_{2s_i}} $, $\epsilon_i$ is the polarization tensor of $i$-th massive particle with spin-$s_i$ and its quantum number under massive little group (LG) $SU(2)_i$ and Lorentz subgroup $SU(2)_r$ is $(2s_i+1,2s_i+1)$, the bracket $\{I_1,\ldots , I_{2s_i} \}$ means these $2s_i$ massive LG indices are totally symmetric, and massive little group neutral structure (MLGNS) $G$ is only charged under massless LGs so it is the function of massless left-handed (right-handed) spinor $\ket{j}$ ($|j]$) and massive momentum $p_{i} \equiv \ket{i^I} [i_I|$~\cite{Arkani-Hamed:2017jhn}. So, to construct the complete set of independent amplitude bases $\mathcal{M}^I_{m,n}$, we can first construct the complete set of independent bases of structure $\mathcal{A}^{\{I\}}$ and then construct the corresponding complete set of independent $G^{\{\dot{\alpha}\}}$ bases. Finally the complete set of amplitude bases can be obtained by contracting each MLGTS basis with its partner MLGNS bases. In~\cite{Dong:2021yak}, it was proven that the amplitude bases constructed in this way are independent without EOM and IBP redundancy.

Since $\epsilon_i$ is in the representation $(\mathbf{2s_i+1})$ of $SU(2)_r$, any MLGTS $\mathcal{A}^{\{I\}}$ must belong to the $SU(2)_r$  reducible representation $\otimes_{i=1}^m (\mathbf{2s_i +1})$ of the $m$ polarizations ($\epsilon_i$). It means that the complete basis of $\mathcal{A}^{\{I\}}$ is all irreducible representations of the outer product of $\otimes_{i=1}^m (\mathbf{2s_i +1})$. Based on Littlewood-Richardson Rule, Young diagram (YD) can systematically find all these irreducible representations. Since the YD of $SU(2)_r$ group reflects the contraction pattern of $SU(2)_r$ indices, the Lorentz structures of these MLGTS bases can be easily read off from the $SU(2)_r$ YDs. Since MLGTS bases are the holomorphic function of right-handed spinors, they can not be EOM and IPB redundant. So MLGTS bases constructed in this way must be independent and complete.

The MLGNS $G$ can suffer from EOM and IBP redundancy. Since EOM of massless spinor is trivial ($p_j |j]=0$ and $p_j \ket{j}=0$), to get rid of EOM redundancy, we can first construct the massless limit of $G$ basis, which is equal to that all the massive momentums in $G$ basis go to the massless limit,
\bea
g(|j],|j\rangle,|i]\langle i|) = G(|j],|j\rangle,p_i)|_{p_i\rightarrow|i]\langle i|},
\eea
where we denote the massless limit of massive momentum $p_{i} = \ket{i^I} [i_I|$ as $|i]\langle i|$. To remove the IBP redundancy in $g$, following the theory proposed in~\cite{Henning:2019enq}, the $N$ right-handed (left-handed) massless spinors of external momentums are embedded into the (anti-) fundamental representation of global symmetry $U(N)$. So the quantum numbers of the $N$ massless spinors under $U(N) \otimes SU(2)_l \otimes SU(2)_r$ are
\bea
\tilde{\lambda}_{\dot \alpha}^k \equiv |k]_{\dot \alpha} &\in&  (N,1,2), \nonumber \\
\lambda_{ \alpha}^k \equiv \ket{k}_\alpha &\in& (\bar{N},2,1), \; k=1,\ldots, N.
\eea
Since the spinors $\tilde{\lambda}_{\dot \alpha}^k$ ($\lambda_{ \alpha}^k$) only take the indices of two groups, the $U(N)$ and Lorentz representations of these spinor polynomials must be correlated. For example, if $g(\tilde{\lambda})$ is the holomorphic function of $\tilde{\lambda}$ and furnishes a representation of $U(N)$, the shape of its $SU(2)_r$ and $U(N)$ YD are always the same. It indicates that $U(N)$ YD of $g(\tilde{\lambda})$ at most has two boxes in each column. If $g(\tilde{\lambda})$ contains $R=r_1+r_2$ right-handed $\tilde{\lambda}$s and is in the ($\bf r_1 -r_2 +1$) representation of $SU(2)_r$, the $U(N)$ YD shape of $g(\tilde{\lambda})$ is $[r_1,r_2]$ ($[r_1,r_2,\cdots r_n]$ means that the YD has $n$ lines and its $j$-th line contain $r_j$ boxes), which can be represented by  following YD
\eq{ \label{eq:leftYD}
    g_{[r_1,r_2]}(\tilde{\lambda})
    =\Yvcentermath1 \renewcommand\arraystretch{0.05} \setlength\arraycolsep{0.2pt}
    \begin{array}{c} \overbrace{\yng(1)\cdots\yng(1)\cdots\yng(1)}^{r_1}\\
    \underbrace{\yng(1)\cdots\yng(1)}_{r_2} {\color{white} \ \ \ \ \ \ \ \,} \end{array}\,.}
 If $SU(2)_l$ singlet $g(\lambda)$ is the holomorphic function of $L$ left-handed $\lambda$s and furnishes a representation of $U(N)$, its $U(N)$ YD consists of $L/2$ boxes in each row and $N-2$ boxes in each column, $[l_1, l_2, \cdots, l_{N-2}]$ with $l_1=l_2 = \cdots =l_{N-2} =L/2$,
\bea \label{eq:rightYD}
    g(\lambda)
    =\Yvcentermath1\underbrace{  \blue{  \begin{array}{ccc}
    \yng(1,1)& \cdots & \yng(1,1) \\
    \vdots &  \ddots & \vdots \\
    \yng(1,1)& \cdots & \yng(1,1)
     \end{array}}  }_{L/2}
\left. \begin{array}{c}
     \\
     \\
     \\
 \end{array}  \!\!\!\!\right \} N-2\,.\!\!\!\!\!\!\!\!\!\!
\eea
Notice that in order to distinguish $\lambda$ from $\tilde{\lambda}$ the boxes in the $U(N)$ YD associated with left-handed spinors are always in blue color. In above two case, the polynomials $g(\tilde{\lambda})$ and $g(\lambda)$ are independent and free of IBP redundancy (corresponds to momentum conservation condition). As discussed in~\cite{Henning:2019enq, Dong:2021yak}, the the non-holomorphic  $g(\lambda, \tilde{\lambda})$ bases, which contain $L$ left-handed spinors $\lambda$s and $R=r_1+r_2$ right-handed spinors $\tilde{\lambda}$s and are in the ($\bf 1,\, r_1-r_2+1$) representation of $SU(2)_l \otimes SU(2)_r$, should furnish the $U(N)$ representation obtained by just gluing the two YDs in Eq~(\ref{eq:leftYD}) and~(\ref{eq:rightYD}) together without changing their shapes,
\bea \label{eq:YD}
    g(\tilde{\lambda},\lambda)
    =\Yvcentermath1 N-2 \left \{ \begin{array}{c}
     \\
     \\
     \\
 \end{array}  \right.\!\!\!\!\!
 \underbrace{  \blue{  \begin{array}{ccc}
    \yng(1,1)& \cdots & \yng(1,1) \\
    \vdots &  \ddots & \vdots \\
    \yng(1,1)& \cdots & \yng(1,1)
    \end{array}}  }_{L/2} \! \! \!
    \renewcommand\arraystretch{0.05} \setlength\arraycolsep{0.2pt}\Yvcentermath1\begin{array}{c} \overbrace{\yng(1)\cdots\yng(1)\cdots\yng(1)}^{r_1}\\
    \underbrace{\yng(1)\cdots\yng(1)}_{r_2} {\color{white} \ \ \ \ \ \ \ \,}
    \\ \\ \\ \\ \\ \\ \\ \\ \\ \\ \\ \\ \\ \\ \\  \\ \\ \\ \\ \\ \\ \\ \\ \\ \\ \\ \\ \\ \\ \\ \\ \\ \\ \\
    \\ \\ \\ \\ \\ \\ \\ \\  \\ \\ \\
    \\ \\ \\ \\ \\ \\ \\ \\ \\ \\ \\ \\ \\ \\ \\  \\ \\ \\ \\ \\ \\ \\ \\ \\ \\ \\ \\ \\ \\ \\  \\ \\
    \end{array}.
    \eea
Then the expression of a $g$ basis can be read off from an SSYT of this kind of $U(N)$ YD (more details can be found in~\cite{Dong:2021yak}). The polynomials in other kinds of $U(N)$ representations are IBP redundant, containing an overall total momentum factor $(\sum^N_{i=1}p_i)$.  Finally, massive $G$ bases can be got by just restoring the LG indices of the massless limit spinors in $g$, $|i], \ket{i}  \to |i^I], \ket{i_I}$, and choosing any contraction pattern of these massive LG indices. Notice that two $G$ bases with different massive LG indices contraction patterns are equivalent (see proof in~\cite{Dong:2021yak}). So massive $G$ bases are one-to-one correspondence to massless $g$ bases and thus are independent and complete.

After constructing the MLGTS $\{\mathcal{A}\}$ and MLGNS $\{G\}$ bases, a complete set of independent amplitude bases $\{\mathcal{A}\cdot G\}$ can be obtained by contracting bare $SU(2)_r$ indices of $\mathcal{A}$ bases with these of their partner $G$ bases.

\subsection{Simplified Amplitude Bases}

However, the SSYT's horizontal permutations make the bases in $\mathcal{A}$ and $G$ very long polynomials. In order to efficiently decompose any polynomial of spinor products into a complete set of $\{\mathcal{A}\cdot G\}$ bases, we should first simplify $\{\mathcal{A}\cdot G\}$ amplitude bases to make each basis be a monomial of spinor products. We find that a set of spinor monomials can be read off from $SU(2)_r$ YDs of a complete set of $\mathcal{A}$ bases without considering the horizontal permutation symmetry (HPS) in these YDs. Moreover, these monomials are independent of each other because of the Fock condition.$^1$ 
\footnotetext[1]{
For general $\mathcal{A}_{\{\dot{\alpha}\}}$ in Eq.~(\ref{eq:AG}), assume it has $n$ bare indices, its horizontal permutation term are not Semi-standard. Using the Fock condition, which is equivalent to the Schouten Identity in the spinor calculation, we can convert $\mathcal{A}$ to $\mathcal{B}$,
$\mathcal{A}_{\{\dot{\alpha}_1,\dots,\dot{\alpha}_n\}}
=\mathcal{B}_{\dot{\alpha}_1,\dots,\dot{\alpha}_n}+\epsilon_{\dot{\alpha}_i\dot{\alpha}_j}
\mathcal{B}_{\dots,\dot{\alpha}_{i-1},\dot{\alpha}_{i+1}
\dots\dot{\alpha}_{j-1},\dot{\alpha}_{j+1}\dots}+\cdots$
Since all polynomial $\mathcal{A}$ could be decomposed into $\{\mathcal{B}\}$, $\{\mathcal{A}\}\subset\{\mathcal{B}\}$; and they have the same number of bases (SSYT), $rank\{\mathcal{A}\}=rank\{\mathcal{B}\}$. Then we can say that these two sets are equivalent, $\{\mathcal{A}\}=\{\mathcal{B}\}$.}
Then we can get the simplified MLGTS bases, called $\mathcal{B}$ bases, which is equivalent to $\mathcal{A}$ basis. Following the same logic, we can also get a complete set of monomials from the $U(N)$ SSYTs of $G$ bases, which is also equivalent to $G$ bases and is called $H$ bases. After contracting the $\mathcal{B}$ bases with the corresponding $H$ bases, the complete set of simplified amplitude basis $\{\mathcal{B}\cdot H\}$ can be obtained (the $SU(2)_r$ Lorentz index contraction convention between $\mathcal{B}$ and $H$ can be fixed in the following discussion).

We find that different from $\{\mathcal{A}\cdot G\}$ basis construction, $\{\mathcal{B}\cdot H\}$ can be easily constructed from the enlarged SSYTs. For the SSYT of a $g$ basis without HPS (see Eq.~(\ref{eq:YD})), the blue sub-SSYT and white sub-SSYT correspond to two monomials, respectively holomorphic function of left-handed spinors and right-handed spinors and their product gives the $H$ basis. Since the shape of white sub-SSYT of $g$ is the same as its $SU(2)_r$ YD,  we can equivalently treat a white sub-SSTY of $U(N)$ as a $SU(2)_r$ YD. So one of $SU(2)_r$ indices contraction patterns between $H$ and $\mathcal{B}$ can be obtained by gluing the white sub-SSYT of $H$ with the $SU(2)_r$ YD of $\mathcal{B}$ via counterclockwise rotating YD of $\mathcal{B}$ by $180\degree$. After gluing, we find that a $\{\mathcal{B}\cdot H\}$ basis corresponds to an enlarged Young Tableau (YT).  In order to distinguish the $H$ part from the $\mathcal{B}$ part in this enlarged YT, we require the boxes representing spinor $|i^I]$ in $\mathcal{B}$ to be labeled by $i^\prime$ while the numbers filling in the boxes associated with $H$ do not take prime superscribe, ranging from $1$ to $N$. Since the sub-YTs associated with $\mathcal{B}$ are rotated by $180\degree$ in order to glue with $H$ SSYTs, in order to make the enlarged YT be `SSYT', we define the size order of the numbers filled in the enlarged YT as
$1<\cdots<N<m'<\cdots<2'<1'$. Since the $i^\prime$-boxes in the enlarged SSYT are only associated with right-handed spinors, the enlarged SSYTs of $\{\mathcal{B}\cdot H\}$ bases should not have blue boxes filled in $i^\prime$.
Conversely, we can easily find that the enlarged SSYTs without HPS satisfying this condition one-to-one correspond to $\mathcal{B}\cdot H$ bases.

As we said before, to get a complete set of amplitude bases with the lowest dimension (a complete set of bases with the lowest dimension means that EOM can not further reduce the dimension of the bases in it), we should first find the complete but redundant bases, which contain all the lowest dimension bases. Then decompose them into $\{\mathcal{B}\cdot H\}$ bases to pick up the independent and lowest dimension monomials as amplitude bases. For the convenience of this decomposition, we re-number the external legs: the $m$ massive legs are labeled by number $\{1, n+2, \cdots, N\}$, and the $n$ massless legs are labeled by number $\{2, \cdots, n+1\}$ according to their spins in descending order.
As before, we define the size order of the numbers filled in the enlarged YT as
\bea \label{eq:order}
    1<\cdots<N<N'<\cdots<(n+2)'<1',
\eea
and thus each enlarged SSYT without blue boxes filled in number-$i^\prime$ and without HPS still one-to-one corresponds to a $\mathcal{B}.H$ basis. Moreover, we will see that any polynomial of spinor products can be decomposed into the $\mathcal{B}.H$ bases constructed from this kind of enlarged SSYTs.

\subsection{General Property of $\{\mathcal{B}\cdot H\}$ Bases and An Example}
\label{sec:BHshape}

The complete set of $\{\mathcal{B}\cdot H\}$ bases can be constructed by finding all the enlarged SSYTs satisfying:
\begin{itemize}
    \item[$\bullet$] Fill YD $[(L+R)/2, (L+R)/2, \left(L/2\right)^{N-4}]$ with $L/2$ number-$i$ and $2s_i$ number-$i'$ for massive particle-$i$ ($i=\{1,n+2,\cdots,N\}$); and $(L/2+2h_j)$ number-$j$ for massless particle-$j$ ($j=\{2,\cdots,n+1\}$).
    \item[$\bullet$] The number-$i'$  can only appear in the white area corresponding to the right-handed spinors in polarization tensors.
\end{itemize}
For a specific $\mathcal{B}.H$ basis with dim-$D$, its SSYT shape should be constrained by the following conditions:
\bea\label{eq:YDofBH}
    R&=&D-N+\sum s_i+\sum h_j \,, \nonumber \\
    L&=& D-N-\sum s_i-\sum h_j\,.
\eea
Here the dimension $D$ of a $\mathcal{B}.H$ basis is defined to be the dimension of its corresponding operator, $D\equiv[\mathcal{B}.H] +N$, where $[\mathcal{B}.H]$ is dimension of $\mathcal{B}.H$ amplitude  and $N$ is external leg number. $\sum s_i$ and $\sum h_j$ are the sums of all massive particle spin and massless particle helicity, respectively.

We take interactions of four massive fields, fermion-fermion-scalar-scalar ($\psi_1\psi_2\phi_3\phi_4$), as an example to explain how to systematically construct a complete set of their $\{\mathcal{B}.H\}$ bases at a given dimension $D$ through enlarged SSYTs.

The polarization tensors of $\psi_1$ and $\psi_2$ are just spinor $|1^I]$ and $|2^J]$ so their enlarged SSYTs should contain two white boxes filled with numbers $1^\prime$ and $2^\prime$ representing their polarization tensors.
Besides the two polarization tensors, $\{\mathcal{B}.H\}$ basis also can contain any number of massive momentums $p_i$. It can contain at least zero momentums, corresponding to $D=5$ bases. According to the conditions in  Eq.~(\ref{eq:YDofBH}), we get $R=2$ and $L=0$ for $D=5$. So, based on the above two properties of $\{\mathcal{B}.H\}$ SSYT, its SSYT is in the shape of $[1,1]$ and is only filled with number $\{2',1' \}$. Fill in this YD with the number $\{2',1'\}$ in the semi-standard pattern according to the number size defined in Eq.~(\ref{eq:order}), and we only get one SSYT,
\bea
    \{\mathcal{B}\cdot H\}^{D=5}&=\Bigg\{ \begin{array}{c} \begin{Young}
    $\,2'$\cr
    $\,1'$\cr
    \end{Young} \end{array}\Bigg\}= \{[1^I2^J]\}.
\eea
Then for higher dimension bases, it may contain two momentums, and its dimension $D=7$. Similarly, we can find that $R=4$ and $L=2$. So the YD shape is $[3, 3]$, and the number-$i$ and $i^\prime$ only appear once. We should fill the number $\{1,2,3,4,2',1' \}$ in this YD in the semi-standard pattern and get five bases at $D=7$:
\eq{
   & \{\mathcal{B}\cdot H\}^{D=7}=\Bigg\{
    \begin{array}{c}
    \blue{\begin{Young} $1$\cr
    $4$\cr \end{Young}}
    \begin{Young} $2$&$3$\cr
    $\,2'$&$\,1'$\cr \end{Young}
    \end{array}\,,\,
    \begin{array}{c}
    \blue{\begin{Young} $1$\cr
    $3$\cr \end{Young}}
    \begin{Young} $2$&$4$\cr
    $\,2'$&$\,1'$\cr \end{Young}
    \end{array}\,,\,
    \begin{array}{c}
    \blue{\begin{Young} $1$\cr
    $3$\cr \end{Young}}
    \begin{Young} $2$&$\,2'$\cr
    $4$&$\,1'$\cr \end{Young}
    \end{array}\,,\\
    &\quad\quad\,\begin{array}{c}
    \blue{\begin{Young} $1$\cr
    $2$\cr \end{Young}}
    \begin{Young} $3$&$4$\cr
    $\,2'$&$\,1'$\cr \end{Young}
    \end{array}\,,\,
    \begin{array}{c}
    \blue{\begin{Young} $1$\cr
    $2$\cr \end{Young}}
    \begin{Young} $3$&$\,2'$\cr
    $4$&$\,1'$\cr \end{Young}
    \end{array}
    \Bigg\}\\
    =&\{[1^I3 22^J], [1^I422^J], s_{24}[1^I2^J], [1^I432^J],s_{34}[1^I2^J]\}\,,}
where $s_{ij}\equiv p_i\cdot p_j$ is the Mandelstam variable. Now we get the complete set of $\{\mathcal{B}.H\}$ bases at $D=5$ and $7$. Following the same procedure, the complete set of bases at any higher $D$ can be systematically obtained.




\section{Decomposition}
\label{sec:decomposition}
In this section, we will discuss how to decompose any polynomial into the $\{\mathcal{B}\cdot H\}$ basis systematically.
Since the massive polarization tensors of $\{\mathcal{B}\cdot H\}$ basis are the holomorphic functions of right-handed spinors, any left-handed spinors in the massive polarization tensors of a polynomial should be replaced by right-handed spinors through EOM ($\ket{i^I}_{\alpha} =p_{i \alpha \dot \alpha } |i^I]^{\dot \alpha}/m_i$) for the decompostion. Since $\{\mathcal{B}\cdot H\}$ bases are constructed from enlarged SSYTs without HPS, the momentum $p_{1 \alpha \dot \alpha}$ and Lorentz scalar $\vev{2_I3_J} [3^K 2^L]$ can not appear in semi-standard $H$ bases.$^2$
\footnotetext[2]{
The Schouten Identity guarantees that the interior of the blue or white part is semi-standard. Now the only non-semi-standard place is where blue and white boxes meet. Since particle-$1$ is massive, there are a total of $L/2$ numbers-$1$ that need to be filled in the Young Tableau, and they just fill all the blue boxes in the first row. So after conjugating the blue boxes, there will be no $|1^I]\langle1_I|$ in  $\mathcal{B}\cdot H$.
In the absence of $p_1$, the only possible non-semi-standard Young tableau where blue and white meet is $\Yvcentermath1 \blue{ \young(1,4)} \young(2,3)$, which corresponds to polynomials with a factor $\vev{23}[23]$ in massless limit.}
So to decompose a polynomial into $\{\mathcal{B}\cdot H\}$ bases, all the $p_{1 \alpha \dot \alpha}$ and $\vev{2_I3_J} [3^K 2^L]$ in it should be eliminated by some identities (see below). After decomposition, each generated spinor monomial can be mapped to an enlarged SSYT, which means this polynomial is correctly decomposed into $\{\mathcal{B}\cdot H\}$ bases. Otherwise, do not stop using Schouten identity until the polynomial is converted into the correct one that can be mapped into the enlarged SSYTs (this can always be realized).

\begin{figure}
\includegraphics[width=4.2cm]{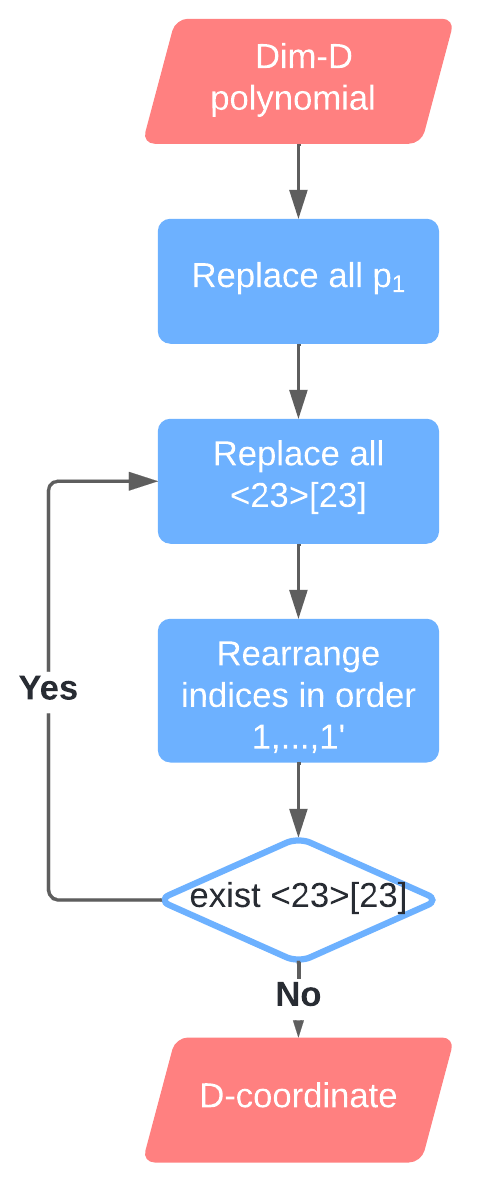}
\caption{Decomposition procedures of any polynomial into $\{\mathcal{B}\cdot H\}$ bases}
\label{fig:BHdecompose}
\end{figure}

In the following, we summarize how to systematically do the decomposition, which is also shown in Fig.~\ref{fig:BHdecompose}. Note that all the polarization's spinors $|i^I\rangle_{\alpha}$ in the input polynomial are replaced with $p_{i \alpha \dot \alpha } |i^I]^{\dot \alpha}/m_i$.
\begin{itemize}
\item{\bf Step-1:}   using momentum conservation, replace all the momentum $p_{1\alpha \dot \alpha } =\ket{1^I}_\alpha [1_I|_{\dot \alpha}$ by $-\sum_{k=2}^Np_k$ in this polynomial and then simplify it by EOMs,
\eq{
    \quad\quad(p_i)_{\alpha\dot{\alpha}}(p_i)^{\dot{\alpha}}_\beta = m^2_i \epsilon_{\alpha\beta}\,&,\ 
    (p_i)_{\alpha\dot{\alpha}}(p_i)^{\alpha}_{\dot{\beta}} = m^2_i \epsilon_{\dot{\alpha}\dot{\beta}}\,, \\
    \vev{jj} = &[jj] =0\,.
}
To guarantee the polarization tensor is always $|i^I]^{2s_i}$, we do not apply EOM $p_i|i^I]= m_i |i^I \rangle$ to the polarization's spinor. If some terms get an overall factor $m^2_i$ from EOMs, these terms should be discarded (we will see that these terms do not affect finding a set of lowest dimension bases).

\item{\bf Step-2:} If one monomial contains $\vev{2^I3^J} [3_J 2_I]$, directly replace it by the identity
\bea \label{eq:p2p3}
     p_2\cdot p_3 &=&  -\frac{m_2^2+m_3^2-m_1^2}{2} -\frac{(\sum_{k=4}^Np_k)^2}{2} \nonumber  \\
    &-&  (p_2+p_3)\cdot \sum_{k=4}^Np_k.
\eea
Suppose one monomial contains $\langle2_I3_J\rangle$ and $[2^K3^L]$ simultaneously, and their LG indices are not bare (these spinors come from momentums $p_2$ and $p_3$). In that case, we use Schouten identity to adjust the LG indices of $\vev{2_I3_J}$ and the other two spinors, $\ket{2}_K$ and $\ket{3}_L$, in this monomial to generate factor $p_2.p_3$. However, in practice, we can directly exchange the LG indices of $\vev{2_I3_J}$ with the spinor $\ket{2}_K$ and $\ket{3}_L$ (if particle-2 or 3 is massless, ignore this step). Since each time we use Schouten identity to adjust the LG index, an additional term with mass factor will be generated, which we should discard. The following is the proof,
\bea \label{eq:LGdiff}
    &&\langle 2_I3_J\rangle\langle 2_Kx\rangle\langle 3_Ly\rangle
    [2^K3^L]  \nonumber \\
    &=&\langle 2_K3_J\rangle\langle 2_Ix\rangle\langle 3_Ly\rangle
    [2^K3^L]
    +\mathcal{O}(m_2^2) \nonumber \\
    &=&\langle 2_K3_L\rangle\langle 2_Ix\rangle\langle 3_Jy\rangle
    [2^K3^L]
    +\mathcal{O}(m_2^2)+\mathcal{O}(m_3^2).
\eea
 Then we can get  $p_2\cdot p_3$ factor and replace it by Eq.~(\ref{eq:p2p3}).

\item{\bf Step-3:}
Since $\{\mathcal{B}\cdot H\}$ basis is constructed based on the SSYT, its spinor contractions are arranged in the order of Eq.~(\ref{eq:order}) (called semi-standard). Generally, after step-$2$, the generated polynomial is not semi-standard. So in order to be decomposed into $\{\mathcal{B}\cdot H\}$ bases, the spinor contraction pattern in this polynomial should be adjusted by Schouten identity to become semi-standard. There is only one kind of spinor contraction pattern that is not semi-standard. That is, if $i<k<l<j$, $[ij][kl]$ ($\vev{ij}\vev{kl}$) is not semi-standard. We can easily convert it into the semi-standard pattern as
\bea
    [ij][kl]= [il][kj]-[ik][lj].
\eea
The two terms on the right side of this equation can be mapped into the sub-SSYT, which shows that any non-semi-standard polynomial can be converted into the combination of semi-standard monomials by Schouten identity.

\item{\bf Step-4:}  Repeat step-$2$ and -$3$ until there are no $p_2.p_3$ factors in the generated polynomial, and each term in it is the semi-standard monomial, which is $\{\mathcal{B}\cdot H\}$ base.
\end{itemize}
Following the above four steps, we can systematically decompose any monomial into $\{\mathcal{B}\cdot H\}$ bases and get its coordinate in $\{\mathcal{B}\cdot H\}$ base space.

\section{Over Redundant $\{\mathcal{C}\cdot F\}$ basis}
\label{sec:CFbase}
Now we know how to systematically decompose any polynomial into $\{\mathcal{B}\cdot H\}$ bases and determine their independence. The only problem is systematically constructing a complete but redundant set of amplitude bases that always contains all lowest dimensional amplitude bases. After constructing such a basis set, we can decompose its bases into $\{\mathcal{B}\cdot H\}$ bases in the ascending order of dimension and eliminate the linear correlation bases according to their coordinates. Finally, we get a complete set of lowest dimensional amplitude bases in the redundant basis set. This section will discuss how to construct this kind of basis set systematically.

\subsection{Lowest Dimension Basis $\{\mathcal{C}\cdot F\}$}
\label{subsec:CFbase}
In this subsection, we briefly discuss why the simplified $\{\mathcal{B}\cdot H\}$ bases or $\{\mathcal{A}\cdot G\}$ bases can not be directly mapped into the physical operator bases. The complete set of physical operator bases always refers to the basis set with the lowest dimension. EOM or other identities can not further reduce the dimension of the bases.
Since the polarization tensors of $\{\mathcal{B}\cdot H\}$ bases are always the holomorphic function of massive right-handed spinors, they can not be mapped into the operator bases whose polarization tensors contain left-handed spinors. If we replace the left-handed spinors in polarizations through EOM $\ket{i^I} = p_i \antiket{i^I}/m_i$, the operator basis is the linear combination of the $\{\mathcal{B}\cdot H\}$ bases with higher dimensions.
We can take four massive particle vertex fermion-fermion-scalar-scalar ($\psi_1\psi_2\phi_3\phi_4$) as an example. Obviously, it has two dim-5 operator bases $\left \{\bar{\psi}_{1L}\psi_{2R} \phi_3 \phi_4, \bar{\psi}_{1R}\psi_{2L} \phi_3 \phi_4 \right \}$, corresponding to the lowest dimensional amplitude bases $\left \{ [1^I2^J], \langle1^I2^J\rangle \right\}$ respectively. As said above, $\langle1^I2^J\rangle$ base does not exist in $\{\mathcal{B}\cdot H\}$ set, which can be expressed as the combination of $\{\mathcal{B}\cdot H\}$ bases,
\bea
    \langle1^I2^J\rangle=\frac{m_2[1^I2^J]}{m_1}
    +\frac{[1^I322^J]}{m_1m_2}+\frac{[1^I422^J]}{m_1m_2}.
\eea
Obviously, the last two  $\{\mathcal{B}\cdot H\}$ bases at the right side of this equation have higher dimensions than $\langle1^I2^J\rangle$ base.


In order to find the lowest dimensional operator bases, we need to know how the amplitude bases are mapped into operator bases and the correlation between the dimension of operator bases and amplitude bases. Generally, the maps between these two kinds of bases follow the rules,
\bea \label{eq:matching}
&&\phi_i \leftrightarrow {\bf1} \,, \;   \psi_{iL} \leftrightarrow \ket{i^I}\,, \; \psi_{iR} \leftrightarrow \antiket{i^I} \,, \; F^+_{i\dot{\alpha}\dot{\beta}}\leftrightarrow \antiket{i^{\{I_1}} \antiket{i^{I_2\}}}\,, \nonumber \\
&&  A_{i\mu} \leftrightarrow \frac{ \antiket{i^{\{I_1}} \langle i^{I_2\}} |}{m}\,, \;
F^-_{i \alpha\beta} \leftrightarrow \ket{i^{\{I_1}} \ket{i^{I_2\}}} \,, \; \partial_i \leftrightarrow  p_{i},
\eea
where $\phi_i$ is scalar, $\psi_{iL,R}$ is left-handed or right-handed massive fermion, $A_{i\mu} $ is massive vector, $F^{\pm}_{i\dot{\alpha}\dot{\beta}} \equiv 1/2( F_{i\mu \nu} \pm i \epsilon_{\mu \nu \rho \sigma} F^{\rho \sigma}_i)$, and $F_{i\mu \nu}$ is field strength of $A_{i\mu} $.
We find that when the operator contains bare vector fields $A_{i\mu}$ (not refer to $A_\mu$ in $F^{\pm}$), the operator dimension $d$ is not equal to the dimension of the corresponding amplitude basis plus the number of external legs $N$, but equal to its amplitude base dimension plus the number of external legs and minus the number of bare vector fields,
\bea \label{eq:operator_dim}
d =[\mathcal{M}] +N -n_A,
\eea
where $[\mathcal{M}]$ is the dimension of amplitude base $\mathcal{M}$, $N$ is the number of external legs, and $n_A$ is the number of $A_{i\mu}$ vector fields.

For a scattering process involving massive external particle-$i$ with spin-$s_i$, we can classify all of its possible amplitude bases according to the particle-$i$'s polarization tensor configuration (PTC),
\bea \label{eq:polarizationConfig}
\epsilon_i^{l_i} \equiv \left( |i^{\{I}\rangle \right)^{l_i}  \left(|i^{I\}}]\right)^{2s_i-l_i}\,,
\eea
where $l_i \in [0,2s_i]$ is the number of left-handed spinors in the polarization tensor, all the LG indices should be totally symmetric. So different sets of $l_i$ value represent different PTCs.
We use the symbol $\{\mathcal{C}\cdot F\}^{\{...,l_i,\, l_{i+1},...\}}_d$ to donate the complete set of dim-$d$ operators with PTC $\{...,\, \epsilon_i^{l_i},\,\epsilon_{i+1}^{l_{i+1}}\,,....\}$. In this kind of set, the basis can also be generally factorized into two parts, similar to $\{\mathcal{B}\cdot H \}$ bases,
\bea
&&\left( \mathcal{C}\cdot F \right)^{\{...,l_i,\, l_{i+1},...\}}= \nonumber \\
& & \mathcal{C}^{\{..,l_i,..\}}\left( |i^{\{I}]^{2s_i -l_i} \right)\cdot
F^{\{..,l_i,..\}}\left( |i^{I\}}\rangle^{l_i}, p_i, |j], |j\rangle \right), \nonumber \\
\eea
where the $\mathcal{C}^{\{..,l_i,..\}}$ part is a holomorphic function of massive right-handed spinors and is required to be a linear function of $\left(2s_i -l_i \right)$ right-handed spinors $|i^I]$, while the $F^{\{..,l_i,..\}}$ part takes both massive and massless LG charges, and the massive LG indices of its $\left(l_i \right)$ left-handed spinors $|i^I \rangle$ are bare.
Similar to $\{\mathcal{B}\cdot H\}$ bases, we can first construct $\mathcal{C}^{\{..,l_i,..\}}\left( |i^{\{I}]^{2s_i -l_i} \right) $ bases and $F^{\{..,l_i,..\}}\left( |i^{I\}}]^{l_i}, p_i, |j], |j\rangle \right)$ bases separately, and then the complete set of $\{\mathcal{C}\cdot F\}^{\{...,l_i,\, l_{i+1},...\}}$ bases can be obtained by contracting these two sets of bases. Same as the construction of $\{\mathcal{B}\}$  and  $\{H\}$ bases, $\mathcal{C}^{\{..,l_i,..\}}\left( |i^{\{I}]^{2s_i -l_i} \right) $ bases can be constructed by $SU(2)_r$ YDs, and $F^{\{..,l_i,..\}}\left( |i^{I}\rangle^{l_i}, p_i, |j], |j\rangle \right)$ bases can be obtained by first constructing their massless limit bases $\{ f^{\{..,l_i,..\}} \}$ via $U(N)$ SSYTs and then restoring the massive spinors' LG indices in $\{ f^{\{..,l_i,..\}} \}$ bases to obtain $F^{\{..,l_i,..\}}\left( |i^{I}\rangle^{l_i}, p_i, |j], |j\rangle \right)$ bases.  The massless limit of an $F^{\{..,l_i,..\}}\left( |i^{I}\rangle^{l_i}, p_i, |j], |j\rangle \right)$ basis is defined to be that the LG indices of all its massive spinors are stripped,
\bea
&&f^{\{..,l_i,..\}}\left( |i\rangle^{l_i}, |i\rangle [i|, |j], |j\rangle \right) \equiv \nonumber \\
&&F^{\{..,l_i,..\}}\left( |i^{I}\rangle^{l_i}, p_i, |j], |j\rangle \right)|_{|i^{I}\rangle \to |i\rangle\,, |i^{I}] \to |i] }\,.
\eea
Notice that $\{ f^{\{..,l_i,..\}} \}$ basis is rigorously not the massless of $\{F^{\{..,l_i,..\} } \}$ basis but is one particular LG component of $\{ F^{\{..,l_i,..\}} \}$ basis. So independent $f^{\{..,l_i,..\}}$ bases one-to-one correspond to independent $F^{\{..,l_i,..\}}$ bases. If some $F^{\{..,l_i,..\}}$ bases correspond to the same $f^{\{..,l_i,..\}}$ basis, these bases must be related by EOM so only one of them is independent. Since the proof of $F^{\{..,l_i,..\}}$ bases independence is the same as $\{ H\}$ bases, we will not discuss it in detail (similar proof can be found in~\cite{Dong:2021yak}).

So in this way, a complete set of  $\{\mathcal{C}\cdot F\}^{\{...,l_i,\, l_{i+1},...\}}$ with one kind of PTC $\{...,l_i,\, l_{i+1},...\}$ can be constructed.$^3$
\footnotetext[3]{
Any polynomial with the polarization in Eq.~(\ref{eq:polarizationConfig}) can be expressed as the inner product of tensors $\mathcal{C}$ and $F$. As long as we construct complete set $\{ \mathcal{C}^{\{..,l_i,..\}} \}$ and $\{ F^{\{..,l_i,..\}} \}$, then $\{\mathcal{C}.F\}$ must be complete (may not be independent).
}
The redundant but complete set of $\{\mathcal{C}\cdot F\}$ bases is the one that contains all the complete sets of bases with different PTCs,
\bea
\{ \mathcal{C}.F \}= \sum_{{\{...\,,\,l_i\,,\,l_{i+1}\,,\, ...\}}}^{...\,,\, 0 \le l_i \le 2s_i\,, ...} \left \{ \{{\mathcal{C}.F} \}^{\{...\,,\,l_i\,,\,l_{i+1}\,,\, ...\}} \right\}.
\eea
Since $\{\mathcal{C}\cdot F\}$ contains all the complete bases for different PTCs, all the lowest dimensional bases are contained in it.
Note that there is redundancy between different sets $\{\mathcal{C}\cdot F\}^{\{...\,,\,l_i\,,\,l_{i+1}\,,\, ...\}}$, while the bases within each set are independent.

Same as $\mathcal{B}\cdot H$ bases, the massless limit bases $\{\mathcal{C}\cdot f\}^{\{...\,,\,l_i\,,\,l_{i+1}\,,\, ...\}}$ with $L^\prime/2$ ($R^\prime/2$) left-handed (right-handed) spinor products can also be constructed through the enlarged SSYTs with only vertical permutation (without HPS) satisfying the following conditions:
\begin{itemize}
    \item[$\bullet$] Fill YD $[(L'+R')/2\,,\,  (L'+R')/2\,,\,  \left(L'/2 \right)^{N-4}]$ with $(L^\prime/2-l_i)$ number-$i$, $(2s_i-l_i)$ number-$i'$ for massive particle-$i$ ($i=\{1,n+2,\cdots,N\}$),  and ($L'/2+2h_j$) number-$j$ for massless particle-$j$ ($j=\{2,\cdots,n+1\}$).
    \item[$\bullet$] Number-$i'$ can only be filled in white boxes corresponding to the right-handed spinors in polarization tensor.
\end{itemize}
For a $\{\mathcal{C}\cdot F\}$ basis with physical dim-$d$,  its SSYT should satisfy the following conditions:
\bea\label{eq:YCF}
    R^\prime &=&d -N +n_A +\sum (s_i -l_i)+\sum h_j \,, \nonumber \\
    L^\prime &=& d-N +n_A -\sum (s_i -l_i) -\sum h_j\,.
\eea
Finally the $\{\mathcal{C}\cdot F\}^{\{...\,,\,l_i\,,\,l_{i+1}\,,\, ...\}}$ bases can be obtained from  $\{\mathcal{C}\cdot f\}^{\{...\,,\,l_i\,,\,l_{i+1}\,,\, ...\}}$ enlarged SSYTs as follows.  We assign the totally symmetric $SU(2)_i$ LG indices $\{I_1,\cdots\!,I_{2s_i}\}$ to the first $l_i$ left-handed spinors $|i\rangle$ in $f$ and $(2s_i-l_i)$ right-handed spinors $|i^\prime]$ in $\mathcal{C}$ (spinor $|i^\prime]$ represents the massive spinor $|i^{I}]$)  and treat them as massive polarization tensor $\epsilon_i^{l_i}$. Then pair all the remaining $|i\rangle$s and $|i]$s into massive momentums in any way, $ |i\rangle|i] \to  p_{i \alpha \dot \alpha} =\, \ket{i^I}_\alpha [i_I|_{\dot \alpha}$.  The procedures to systematically construct $\{\mathcal{C}\cdot F\}$ bases is the same as $\{\mathcal{B} \cdot H\}$ bases. An example to explain $\{\mathcal{C}\cdot F\}$ base construction is shown in subsection~\ref{subsec:example}.



\subsection{Decompose $\{\mathcal{C}\cdot F\}$ into $\{\mathcal{B}\cdot H\}$}
After obtaining the complete $\{\mathcal{C}\cdot F\}$ bases, we can do the decomposition $\{\mathcal{C}\cdot F\} \to \{\mathcal{B}\cdot H\}$ following the procedure in Fig.\ref{fig:BHdecompose}, and then eliminate the linear correlation bases. As said before, when doing the decomposition, all the left-handed spinors $\ket{i^I}$ in $\epsilon_i^{l_i}$ should be replaced by $p_i \antiket{i^I}/m_i$ to convert $\epsilon_i^{l_i}$ to holomorphic tenser $\epsilon_i^{0}$, which results in the dimension of $\{\mathcal{C}\cdot F\}$ amplitude bases being increased.
The generated amplitude bases, denoted as $\{\mathcal{C}^\prime\cdot F^\prime\}$ (called holomorphic bases), are equivalent to $\{\mathcal{C}\cdot F\}$, and the $\mathcal{C}'$ and $F'$ part is defined the same as $\mathcal{B}$ and $H$. After converting $\{\epsilon_i^{l_i}\}$ to $\{\epsilon_i^{0}\}$, we can define the operator dimension of $\{ \mathcal{C}^\prime\cdot F^\prime  \}^{\{...\,,\,l_i\,,\,l_{i+1}\,,\, ...\}} _d$ as the fake dimension $D'$ of $\{ \mathcal{C}\cdot F \}^{\{...\,,\,l_i\,,\,l_{i+1}\,,\, ...\}} _d$ bases,
\bea
D'=d + \sum l_i +n_A,
\eea
where $d$ is the operator basis dimension defined in Eq.~(\ref{eq:operator_dim}) and $\sum l_i$ is the total number of left-handed spinors in the polarization tensors (fake dimension $D'$ of a $\{\mathcal{B}\cdot H\}$ base equals to its operator dimension $D$).

We can easily find that a $\{ \mathcal{C}\cdot F \}^{\{...\,,\,l_i\,,\,l_{i+1}\,,\, ...\}} _d$ basis with fake dimension $D'=D_{\mathcal{C}\cdot F}$ can be decomposed into the $\{\mathcal{B}\cdot H\}$ bases with the highest fake dimension equal to  $D_{\mathcal{C}\cdot F}$,
\bea \label{eq:CmapBH}
&& \{\mathcal{C}\cdot F\}^{\{l\}}_d\rightarrow  \{\mathcal{C}^\prime \cdot F^\prime \}^{\{l\}}_d \nonumber \\
 &&\to \{\mathcal{B}\cdot H\}^{D=D_{\mathcal{C}\cdot F}}+m^2\{\mathcal{B}\cdot H\}^{D_{\mathcal{C}\cdot F}-2}+\cdots.
\eea
So decompose $\{\mathcal{C}\cdot F\}_d$ into $\{\mathcal{B}\cdot H\}$ bases in the ascending order of $d$, and remove all the linear correlation bases according to their coordinates in $\{\mathcal{B}\cdot H\}$ bases. Finally, we obtain a complete set of amplitude bases with the lowest dimension, denoted as $\{\mathcal{O}^{phy} \}$.

The above full decomposition of $\{\mathcal{C}\cdot F\}_d$ bases is very inefficient. We find that their coordinates in the $\{\mathcal{B}\cdot H\} ^{D=D_{\mathcal{C}\cdot F}}$ basis space is enough to pick up the independent $\{\mathcal{O}^{phy}\}$ bases (proof is shown in App.~\ref{app:proof}). Therefore, we can discard all terms proportional to mass factors in eq.~(\ref{eq:CmapBH}).
\bea \label{eq:BHdecom1}
 \{\mathcal{C}\cdot F\}^{\{l\}}_d\rightarrow  \{\mathcal{C}^\prime \cdot F^\prime \}^{\{l\}}_d  \to \{\mathcal{B}\cdot H\}^{D=D_{\mathcal{C}\cdot F}}.
\eea
That is to say, any mass factors appear in the decomposition procedures discussed in Sec.~(\ref{sec:decomposition}) should be discarded.

\subsection{Example}
\label{subsec:example}
In this subsection, we explain how to systematically construct $\{\mathcal{C}\cdot F\}$ bases and do the decomposition.
We still take four-point interactions $\psi_1-\psi_2-\phi_3-\phi_4$ as an example.

According to above discussions, since $\psi_{1,2}$ and $\phi_{1,2}$ are fermions and scalars, the range of their polarization parameters $l_i$ is (see Eq.~(\ref{eq:polarizationConfig}))
\bea
l_{1,2} \in [0,1]\,, \,\, l_{3,4} =0\,.
\eea
So there are total four different PTCs.
To obtain all the $d\leq7$ physical operators, we need to construct all the related $\{\mathcal{C}\cdot f\}^{\{l\}}_d$ bases in ascending order of $d\leq7$. According to Eq.~(\ref{eq:YCF}), since $L^\prime$ and $R^\prime$ are even integers, we can determine the allowed PTCs for different dimensions $d\leq7$, and all are listed as follows,
\bea \label{eq:CF_e.g.}
&D'=5:&\; \{\mathcal{C}\cdot F\}^{0000}_5. \nonumber \\
&D'=7:&\; \{\mathcal{C}\cdot F\}^{1100}_5,\{\mathcal{C}\cdot f\}^{0100}_6,
\{\mathcal{C}\cdot F\}^{1000}_6,\{\mathcal{C}\cdot f\}^{0000}_7.\nonumber \\
&D'=9:&\; \{\mathcal{C}\cdot F\}^{1100}_7\,,\cdots
\eea
where the sets in the same line have the same fake dimension-$D'$.
Then, following the procedures in Fig.~\ref{fig:BHdecompose} and Eq.~(\ref{eq:BHdecom1}), decompose the sets in the three lines into $\{\mathcal{B}\cdot H\}^{D=5}$, $\{\mathcal{B}\cdot H\}^{D=7}$, and $\{\mathcal{B}\cdot H\}^{D=9}$ respectively (discard all the mass factors during the decompositions). Here we explicitly show how to construct the above bases at send line ($D=7$) and decompose them. Following Eq.~(\ref{eq:YCF}), we can find that the SSYTs of all these $\{\mathcal{C}\cdot F\}^{\{l\}}_d$ bases are
\eq{
    \{\mathcal{C}\cdot f\}^{1100}_5&=\Bigg\{
    \begin{array}{c}
    \blue{\begin{Young} $3$\cr
    $4$\cr \end{Young}}
    \end{array}
    \Bigg\},
    \\
    \{\mathcal{C}\cdot f\}^{0100}_6&=\Bigg\{
    \begin{array}{c}
    \blue{\begin{Young} $1$\cr
    $4$\cr \end{Young}}
    \begin{Young} $3$\cr
    $\,1'$\cr \end{Young}
    \end{array}\,,\,
    \begin{array}{c}
    \blue{\begin{Young} $1$\cr
    $3$\cr \end{Young}}
    \begin{Young} $4$\cr
    $\,1'$\cr \end{Young}
    \end{array}
    \Bigg\},
    \\
    \{\mathcal{C}\cdot f\}^{1000}_6&=\Bigg\{
    \begin{array}{c}
    \blue{\begin{Young} $2$\cr
    $4$\cr \end{Young}}
    \begin{Young} $3$\cr
    $\,2'$\cr \end{Young}
    \end{array}\,,\,
    \begin{array}{c}
    \blue{\begin{Young} $2$\cr
    $3$\cr \end{Young}}
    \begin{Young} $4$\cr
    $\,2'$\cr \end{Young}
    \end{array}
    \Bigg\},
    \\
    \{\mathcal{C}\cdot f\}^{0000}_7&=\Bigg\{
    \begin{array}{c}
    \blue{\begin{Young} $1$\cr
    $4$\cr \end{Young}}
    \begin{Young} $2$&$3$\cr
    $\,2'$&$\,1'$\cr \end{Young}
    \end{array}\,,\,
    \begin{array}{c}
    \blue{\begin{Young} $1$\cr
    $3$\cr \end{Young}}
    \begin{Young} $2$&$4$\cr
    $\,2'$&$\,1'$\cr \end{Young}
    \end{array}\,,\,
    \begin{array}{c}
    \blue{\begin{Young} $1$\cr
    $3$\cr \end{Young}}
    \begin{Young} $2$&$\,2'$\cr
    $4$&$\,1'$\cr \end{Young}
    \end{array}\,,\\
    &\quad\quad\,\begin{array}{c}
    \blue{\begin{Young} $1$\cr
    $2$\cr \end{Young}}
    \begin{Young} $3$&$4$\cr
    $\,2'$&$\,1'$\cr \end{Young}
    \end{array}\,,\,
    \begin{array}{c}
    \blue{\begin{Young} $1$\cr
    $2$\cr \end{Young}}
    \begin{Young} $3$&$\,2'$\cr
    $4$&$\,1'$\cr \end{Young}
    \end{array}
    \Bigg\}.
}
Read out all the massive amplitude bases according to the discussions in Sec.~\ref{subsec:CFbase} and convert them into holomorphic bases $\{\mathcal{C}^\prime \cdot F^\prime\}^{\{l\}}_d$ by EOMs. And then decompose them into $\{\mathcal{B}\cdot H\}^{D=7}$ in the ascending order of $d$ following the four steps in Sec.~\ref{sec:decomposition}  (see Fig.~\ref{fig:BHdecompose}),
\eq{
   & \langle1^I2^J\rangle \rightarrow\langle12\rangle[11^I][22^J]
    =\langle23\rangle[31^I][22^J]+\langle24\rangle[41^I][22^J]\,,\\
  & \langle2^J3\rangle[31^I] \rightarrow \langle23\rangle[31^I][22^J]
    =\langle23\rangle[31^I][22^J] \,,\\
   & \cancel{\langle42^J\rangle[41^I]} \rightarrow\langle42\rangle[41^I][22^J]
    =-\langle24\rangle[41^I][22^J] \,,\\
    & \quad \quad \quad \quad \quad \quad \quad \quad \quad\cdots\,.}
Here since the decomposition is at the leading order in masses, the massive momentums in the above bases can be treated as massless, and spinor LG indices can be neglected. In contrast, the LG indices from polarization tensors should be kept.
We can see that these three bases are linearly correlated, so we remove the third redundant base with higher $d$. After removing all the linear correlation terms in Eq.~(\ref{eq:CF_e.g.}) and recovering the massive momentums from massless limit, we get the bases with the lowest dimension, which is equivalent to set $\{\mathcal{B}\cdot H\}$,
\eq{
   \{\mathcal{B}\cdot H\}^5 &\rightarrow\{[1^I2^J]\}\,, \\
   \{\mathcal{B}\cdot H\}^7 &\rightarrow\{\langle1^I2^J\rangle,[1^I 32^J\rangle,\langle1^I32^J],s_{24}[1^I2^J],s_{34}[1^I2^J]\}\,, \\
    \{\mathcal{B}\cdot H\}^9 & \rightarrow\{s_{34}\langle1^I2^J\rangle,\langle1^I 342^J\rangle,\cdots\}.
}
With these lowest dimensional bases at dimension $d=5,6,7$  on the right hand side (`$\cdots$' refer to higher dim-$d$ operators), we can directly map them into operator bases following the rules in Eq.~(\ref{eq:matching}).

\section{Identical Particles}
\label{sec:identical}

If the scattering process involves $n$ identical bosons (fermions), the scattering amplitude should be a totally (ant-) symmetric representation of permutation ground $S_n$. Next, we will discuss systematically constructing the complete set of amplitude bases involving identical particles.

The basic idea to construct a complete set of amplitude bases $\{\mathcal{M}_d \}$ at dim-$d$  in the required $S_n$ representation $[R_{n}]$ is that: First, use the Young operator $\mathcal{Y}_{[R_n]}$ of representation $[R_{n}]$ acting on the space of bases $\{\mathcal{M}_d \}$ to get the matrix representation $M_{[R_n]}$ of $\mathcal{Y}_{[R_n]}$. Then, the eigenvectors with non-zero eigenvalues (actually is 1) correspond to the amplitude bases in the $[R_{n}]$ representation. In contrast, the eigenvectors with zero eigenvalues correspond to the bases that vanish under identical particle permutations.

Generally, an amplitude base consists of two parts: the gauge structure ($T$) and Lorentz structure ($\mathcal{D}$),
\bea
    \mathcal{M} = T \times \mathcal{D}\,.
 \eea
So the complete set of amplitude bases can be constructed by combining the complete sets of gauge structure  and  Lorentz structure bases, $\{T\}$ and  $\{\mathcal{D}\}$,
\bea
\{ \mathcal{M} \} = \{T\} \times \{\mathcal{D}\}\,.
\eea
Notice that any Young operator of the permutation group $S_n$ can be expressed as a function of permutation elements $(12)$ and $(1\ldots n)$, so we only need to get their representation matrices in $\{T\}$ and $\{\mathcal{D}\}$ space, $M^T_{(12),(1,\ldots,n)}$ and $M^{\mathcal{D}}_{(12),(1,\ldots,n)}$. So representation matrix $M_{[R_n]}$ is determined by the outer product of matrices $M^T_{(12),(1,\ldots,n)}$ and $M^{\mathcal{D}}_{(12),(1,\ldots,n)}$. For example, the matrix of totally symmetric representation $[3]$ of $S_3$ group can be expressed as
\eq{ \label{eq:B/Fmatrix}
    &M_{\scriptsize\young(123)}=(\mathds{1}+x+y+xy+yx+xyx)/6\,,\\
    &x=M^T_{(12)}\otimes M^{\mathcal{D}}_{(12)},\quad y=M^T_{(123)}\otimes M^{\mathcal{D}}_{(123)},
}
where $M^{a}_{(12),\, (123)}$ is the representation matrix of $(12)$ or $(123)$ in $a \in \{T, \mathcal{D}\}$ space.
In App.~\ref{app:permutate}, we present the systematical method to calculate their representation matrices.

Similarly, if the amplitude bases has different identical bosons (fermions), we only need to multiply the matrix of the totally (anti-)symmetric representation Young operator of each permutation group to get a total matrix and the eigenvectors with non-zero eigenvalues are the bases allowed by identical particle statistic.

Generally, the amplitude bases $\{\mathcal{M}_{id}\}$ satisfying identical particle statistics are polynomials of $\{\mathcal{O}^{phy} \}$. When we map these amplitudes into operator bases through the relation in Eq.~(\ref{eq:matching}), the operator bases are also polynomials. Practically when an amplitude monomial is mapped into an operator, this operator automatically satisfies identical particle statistics (its Feynman rule automatically enforces identical particle permutation symmetries in its operator). So we do not need to map all the components of the $\mathcal{M}_{id}$ basis but one of its independent monomial terms into an operator basis. It is obvious that the operator of this amplitude monomial must be equal to the operator of amplitude $\mathcal{M}_{id}$. So, in this way, we can get a complete set of simplest operator bases equivalent to $\{\mathcal{M}_{id}\}$, and they can be used more conveniently in calculations. In App.~\ref{sec:operator}, we list a complete set of four-vector operator bases at dimension-$4$ and $6$.

\section{conclusion}
\label{sec:conclusion}
EFT of massive fields is widely applied in various fields of physics. However, how to systematically construct the complete set of EFT bases of massive fields is still a long-standing problem. Based on on-shell scattering amplitude, we propose a novel theory to construct the complete set of lowest dimensional amplitude bases at any given dimension for massive fields with any spins. These bases can be directly mapped into physical operator bases without any redundancy.

The massive amplitude bases with the lowest dimension can be constructed through three steps. First, we systematically construct a complete set of massive amplitude bases $\{\mathcal{B}\cdot H \}$ by the enlarged SSYTs without horizontal permutation symmetries, which are constructed by gluing the SSYTs of Lorentz subgroup $SU(2)_r$ and global symmetry $U(N)$. These bases are just simple monomials of spinor products but not the lowest dimensional amplitude bases. Second, since massive amplitude bases can be classified by configurations of massive polarization tensors, we systematically construct a complete but redundant basis set $\{\mathcal{C}\cdot F\}$ that consists of all the complete sets of massive amplitude bases with different polarization configurations. So this redundant set always contains a complete set of lowest dimensional amplitude bases. Finally, since $\{\mathcal{B}\cdot H \}$ bases are complete and independent monomials of spinor products, we can systematically decompose the $\{\mathcal{C}\cdot F\}$ bases into $\{\mathcal{B}\cdot H \}$ bases from low to high dimension and eliminate the linear correlation bases through their coordinates in $\{\mathcal{B}\cdot H \}$ space. After these procedures, we can always obtain a complete set of the lowest dimensional amplitude bases. We also give an example to explain how to get this kind of bases systematically.

The amplitude bases involving identical particles can also be systematically constructed. First, we find the representation matrices of the Young operators, associated with the permutation symmetry representations required by spin statistics, in the amplitude basis space, and then multiply these matrices to get a total matrix, finally the eigenvectors with non-zero eigenvalues of this matrix are the bases satisfying spin statics.

Based on this theory, we write down the Mathematica codes that can automatically construct a complete set of lowest dimensional amplitude bases at a given dimension. We also show the complete sets of all four vectors operator bases at dimension-$4$ and $6$ in App.~\ref{sec:operator} (also the complete bases involving identical particles). A complete set of other kinds of massive operator bases will be presented in the later work~\cite{Donziyu}.

Within this theory, constructing massive EFT is not a problem. Our work provides an efficient tool to study the low energy effects of UV theories at the EWSB phase. The wave function normalization of massive particles at the EWSB phase does not need to be cared about, and a compete sets of three-point and four-point massive EFT bases are enough for phenomenology study generally. So when doing the massive operator matching, we do not need to deal with the high point EFT bases, which can simplify the calculations very much. While in massless EFT, such as SMEFT, many higher point bases involving Higgs doublets always contribute to wave functions of particles, three-point, and four-point interactions at the EWSB phase, which makes EFT calculations complicated. We can also use them to study dark matter interactions with experimental detections and analyze the dark matter signals from different UV models. Massive EFT could have some advantages in various scenarios of physics, and a lot of its exciting applications deserve to be explored in the future.

\section*{Note added}
While our paper was being finalized, Ref.~\cite{DeAngelis:2022qco} appeared, which presents a similar topic. This work uses a graphic method to construct the massive amplitude bases, which is equivalent to the Young Tableaux method used here. Nevertheless, some of the assumptions in Ref.~\cite{DeAngelis:2022qco} are only numerically checked without rigorous proof. On the contrary, our work has a solid mathematical foundation.

\section*{Acknowledgements}
This work is supported by the National Key Research and Development Program of China under Grant No. 2020YFC2201501. T.M. is supported by ``Study in Israel" Fellowship for Outstanding Post-Doctoral Researchers from China and India by PBC of CHE and partially supported by grants from the NSF-BSF (No. 2018683), by the ISF (grant No. 482/20) and by the Azrieli foundation. J.S. is supported by the National Natural Science Foundation of China under Grants No. 12025507, No. 12150015, No.12047503; and is supported by the Strategic Priority Research Program and Key Research Program of Frontier Science of the Chinese Academy of Sciences under Grants No. XDB21010200, No. XDB23010000, and No. ZDBS-LY-7003 and CAS project for Young Scientists in Basic Research YSBR-006.



\onecolumngrid
\newcommand\ptwiddle[1]{\mathord{\mathop{#1}\limits^{\scriptscriptstyle(\sim)}}}

\appendix

\newpage

\section{proof of leading order decomposition}
\label{app:proof}
In this section, we prove that the independence of physical amplitude bases $\{\mathcal{O}^{phy}\}$ with fake dimension $D^\prime =D_{\mathcal{O}^{phy}}$ is determined by their coordinates in the space of $\{\mathcal{B}.H\}^{D=D_{ \mathcal{O}^{phy} }}$ bases.  

For the first case, we suppose that two independent physical bases $\mathcal{O}^{phy}_{1,2}$ have the same PTC and the same fake dimension $D^\prime =D_{\mathcal{O}^{phy}}$  (for simplicity only $l_i$ in their PTC is assumed to be non-zero $\vec{l}= \{...,0,l_i,0...\}$) and their coordinates in the $\{\mathcal{B}\cdot H\}^{D=D_{\mathcal{O}^{phy}}}$ space are assumed to be the same for simplicity. So, after converting the $\mathcal{O}^{phy}_{1,2}$ into holomorphic bases $\mathcal{O}^{phy}_{1,2} \to \mathcal{O}^{\prime phy}_{1,2} = ([i^Ji^I])^{l_i}\mathcal{O}^{phy \{J^{l_i}\}}_{1,2}$ ($J^{l_i}$ represents the LG indices of $l_i$  $\ket{i^J}$s in PTC), we can get the decomposition of the two bases difference,
\bea \label{eq:Odecomposition}
&&([i^Ji^I])^{l_i}\left(\mathcal{O}^{phy \{J^{l_i}\}}_{1} -\mathcal{O}^{phy \{J^{l_i}\}}_{2}\right)
\nonumber \\
&&\to \{\mathcal{B}\cdot H\}^{D=D_{\mathcal{O}^{phy}}-2 }  +\{\mathcal{B}\cdot H\}^{D=D_{\mathcal{O}^{phy}}-4 }+ \cdots\,.
\eea
Since the PTCs at both sides are the same, the decomposition is only determined by the MLGNSs of holomorphic bases  $\mathcal{O}^{\prime phy}_{1,2}$. If the their MLGNSs go to massless limit (massive momentums go to massless limit) and the spinors in PTCs keep intact, above decomposition should become null
\bea \label{eq:Odecomposition1}
&&([i^Ji^I])^{l_i}\left(\mathcal{O}^{phy \{J^{l_i}\}}_{1} -\mathcal{O}^{phy \{J^{l_i}\}}_{2}\right) |_{p_i \to \ket{i} |i]} \nonumber \\
&&=([i i^I])^{l_i} \left( (\mathcal{C}.f)_{\mathcal{O}_1^{phy}}^{\vec{l}} -(\mathcal{C}.f)_{\mathcal{O}_2^{phy}} ^{\vec{l}} \right)   = 0,
\eea
where $(\mathcal{C}.f)_{\mathcal{O}_{1,2}^{phy}}^{\vec{l}}$ are the massless limits of $\mathcal{O}^{phy}_{1,2}$ defined in Sec.~\ref{subsec:CFbase}.
This is because the coordinates in $\{ \mathcal{B}\cdot H\}^{D \le D_{\mathcal{O}^{phy}}-2 }$ space always proportional to mass factor due to fake dimension mismatch at both sides of Eq.~(\ref{eq:Odecomposition}) and the mass factor is only generated from spinor EOMs in $\mathcal{O}^{\prime phy}_{1,2}$ MLGNSs.  Since the identity in Eq.~(\ref{eq:Odecomposition1}) is independent of the spinors in PTCs, these spinors $|i^I]$s can be treated as independent variables. So the only solution is $(\mathcal{C}.f)_{\mathcal{O}_{1}^{phy}}^{\vec{l}} =(\mathcal{C}.f)_{\mathcal{O}_{2}^{phy}}^{\vec{l}}$, which means the two operators $\mathcal{O}_{1,2}^{phy}$ are the same (see the discussions in Sec.~\ref{subsec:CFbase}). It conflicts with our assumption, so we prove that the independence of lowest dimenisonal bases $\{\mathcal{O}^{phy}_{1,2}\}$ with the same PTC is determined by leading decomposition. Above proof can be easily generalized to the case for any number bases.

For the second case, two physical bases $\mathcal{O}^{phy}_{1^\prime,2^\prime}$ have different PTCs but have the same fake dimension $D =D^\prime_{\mathcal{O}^{phy}}$ and their coordinates in the $\{\mathcal{B}\cdot H\}^{D=D^\prime_{\mathcal{O}^{phy}}}$ space are also assumed to be the same for simplicity. For simplicity we assume in their PTCs only $l_i$ and $l_j$ are non-zero respectively, $\vec{l}_{1} =\{0,...,0,l_i,0,...,0\}$ and $\vec{l}_{2} =\{0,...,0,l_j,0,...,0\}$. Following the same logic, we can also get the similar massless limit identity,
\bea
&&\left( ([i^{I^\prime} i^I])^{l_i} \mathcal{O}^{phy \{(I^\prime)^{l_i}\}}_{1^\prime} - ([j^{J^\prime}j^J])^{l_j}\mathcal{O}^{phy \{(J^\prime)^{l_j}\}}_{2^\prime} \right) |_{p_i \to \ket{i} |i]} \nonumber \\
&&=([i i^I])^{l_i} (\mathcal{C}.f)_{\mathcal{O}_1^{phy}}^{\vec{l}_{1} } -([j j^J])^{l_j}(\mathcal{C}.f)_{\mathcal{O}_2^{phy}} ^{\vec{l}_{2}}   = 0.
\eea
As discussed above, since $([i i^I])^{l_i}$ and $([j j^J])^{l_j}$ are two independent LG tensors,
$(\mathcal{C}.f)_{\mathcal{O}_{1}^{phy}}^{\vec{l}_{1}}$ and $(\mathcal{C}.f)_{\mathcal{O}_{2}^{phy}}^{\vec{l}_{2}}$  should contain the tensor factor $([j j^J])^{l_j}$  and  $([i i^I])^{l_i}$ respectively to guarantee that this identity can be satisfied. If so, it means that the massive bases $\mathcal{O}^{phy}_{1^\prime,2^\prime} =(\mathcal{C}.F)_{\mathcal{O}_{1,2}^{phy}}^{\vec{l}_{\mathcal{O}}}$ should contain factor $([j^{J^\prime}j^J])^{l_j}$ ($([i^{I^\prime} i^I])^{l_i}$) . However this factor is proportional to mass so $\mathcal{O}^{phy}_{1^\prime,2^\prime}$ are not the lowest dimension bases, conflicting with  $\mathcal{O}^{phy}_{1^\prime,2^\prime}$ definition. So the independence of two lowest dimensional bases with the same fake dimension and different PTCs is determined by the coordinates in $\{ \mathcal{B}\cdot H\}^{D = D^\prime_{\mathcal{O}^{phy}}}$ space.

Above two proofs can be easily generalized to the case for any number bases.
So in summary the $\{\mathcal{O}^{phy} \}$ bases can obtained just through the leading decomposition of $\{\mathcal{C}\cdot F\}_d$ in Eq.~(\ref{eq:CmapBH}),
 \bea
 \{\mathcal{C}\cdot F\}^{\{l\}}_d\rightarrow  \{\mathcal{C}^\prime \cdot F^\prime \}^{\{l\}}_d  \to \{\mathcal{B}\cdot H\}^{D=D_{\mathcal{C}\cdot F}}.
\eea
Since the coefficients of this leading decomposition are independent of mass factors, any mass factors appear in the decomposition procedures discussed  in Sec.~(\ref{sec:decomposition}) should be discarded.

\section{Representation matrix for $S_n$}
\label{app:permutate}
\subsection{Gauge part}
In this section we explain how to calculate the Yong operator matrix $M^T_{[R_n]}$ of $S_n$ representation $[R_n]$ in gauge structure $\{T \}$ space.
For concreteness, we take $SU(3)_c$ color group as example.

 We suppose that the amplitude bases contain QCD quarks, anti-quarks and gluons.  According to their $SU(3)_c$ quantum numbers, the YDs of their $SU(3)_c$ representation are in the forms,
\eq{
    \psi^a\sim\Yvcentermath1\young(a)
    \quad\quad\quad
    \epsilon^{abc}\bar{\psi}_a\sim\Yvcentermath1\young(b,c)
    \quad\quad\quad
    \epsilon^{abc}\lambda_a^{id}g^i\sim\Yvcentermath1\young(bd,c).
}
where $\lambda^{i}$ refer to the eight Gell-Mann matrices. The complete set of   $SU(3)_c$  $\{T\}$ bases consists of all the Standard Young Tableaus (SYTs) in the shape of $[x\,, x\,, x]$ ($SU(3)_c$ singlet) satisfying above $SU(3)_c$ index permutation symmetries of all $\bar{\psi}_a$s and $g^i$s. In the following, we will discuss how to systematically construct the $\{T\}$ bases and then discuss how to get the matrices $M^{T}_{[R_n]}$.

We can first find a set of $SU(3)_c$ gauge structure bases $\{T^\prime\}$ consists of all the SYTs with shape of $[x,x,x]$, where the length $x$ of each row is determined by the $SU(3)_c$ quantum numbers of external fields. The set of $\{T^\prime\}$ bases contain $\{T\}$ base we need. Then we can use the projection matrix $\mathcal{P}$ (see below) to project out the structures whose indices satisfying the permutation symmetry of all $\bar{\psi}_a$s' and $g^i$s' $SU(3)_c$ indices. So these projected out bases are the $\{T\}$ bases. Next we will list the procedures to get matrices $M^{T}_{[R_n]}$.
\begin{itemize}
\item{Get the representation matrix $M^{T^\prime}_{(12)/(1\dots n)}$ of $S_n$ element $(12)$ or $(1\dots n)$ in $\{T^\prime \}$ space:}\\
 Since there are totally $n_Y\equiv\frac{2(3x)!}{x!(x+1)!(x+2)!}$ SYTs with shape $[x,x,x]$, the number of $\{T^\prime\}$ bases is equal to $n_Y$. The $(n_Y\times n_Y)$ representation matrix $M^{T^\prime}_{(12)}$ of $(12)$ can be obtained by first using permutation element $(12)$ to act on the color indices of $\{T^\prime\}$ base associated with particle-$1$ and -$2$ and then decomposing them into the $n_Y$ SYTs. $M^{T^\prime}_{(1\dots n)}$ can also be obtain in the same way.

\item{Get the projection matrix $\mathcal{P}$ :}\\
    $\mathcal{P}$ is the product of adjoint (anti-fundamental) Young Operators of each external particle $g$ ($\bar{\psi}$) .
     \eq{
    \mathcal{P}\equiv \overbrace{\mathcal{Y}_{\tiny\yng(2,1)}\cdots\mathcal{Y}_{\tiny\yng(2,1)}}^{adjoint}\
    \overbrace{\mathcal{Y}_{\tiny\yng(1,1)}\cdots\mathcal{Y}_{\tiny\yng(1,1)}}^{anti-fund.}\,.}
The amplitude gauge structure  bases $\{T\}$ can be projected out  from the enlarged $\{T^\prime\}$ bases by acting $\mathcal{P}$ on the $n_Y$ SYTs and decomposing them back to the $n_Y$ SYTs. Finally we can get the $(n_Y\times n_Y)$ matrix representation $P$ of $\mathcal{P}$ in $\{T^\prime\}$ space. The eigenvectors with non-zero eigenvalues of $P$ are the $\{T\}$ bases consistent with $SU(3)_c$ quantum number of external legs.
\end{itemize}
The method used in above decompositions is just Fock condition. Using this method, we can turn any Young Tableau into a combination of SYTs.
Finally, we get the $SU(3)_c$ representation matrix for $(12)$ and $(1\cdots n)$ in the space of $\{T \}$ bases ($P$ is to project out $\{T\}$ bases from $\{T^\prime\}$),
\eq{
    M^T_{(12)/(1\cdots n)}=M^{T^\prime}_{(12)/(1\cdots n)}\cdot P\,.
}
Since $M^{T^\prime}$ and $P$ are commutative, we can commutate all $P$ in Eq.~(\ref{eq:B/Fmatrix}) to the most right position, which can simplify the calculation. Finally the  total symmetric representation of  $S_3$ can be expressed as
\eq{
    &M_{\scriptsize\young(123)}=(\mathds{1}+x+y+xy+yx+xyx)(P\otimes\mathds{1}^{\mathcal{D}})/6,\\
    &x=M^{T^\prime}_{(12)}\otimes M^\mathcal{D}_{(12)},\quad y=M^{T^\prime}_{(123)}\otimes M^\mathcal{D}_{(123)}.}

\subsection{Lorentz part}
\begin{figure}
\includegraphics[width=7cm]{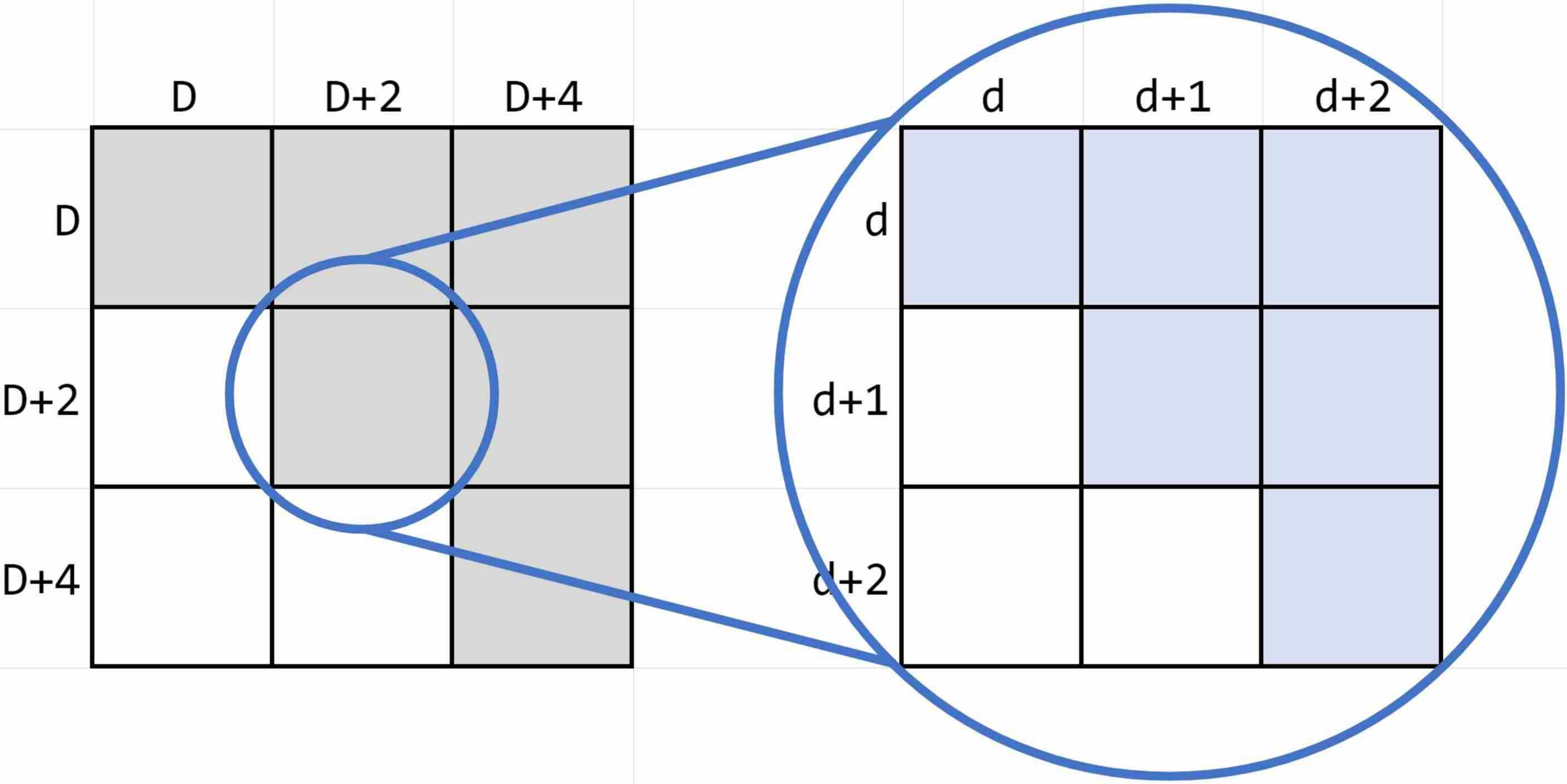}
\caption{Partitioned upper triangular matrix $M^{\mathcal{D}}_{(12)/(1\cdots n)}$}
\label{fig:Dd_matrix}
\end{figure}

To get $M^\mathcal{D}_{(12)}$ in bases $\{\mathcal{O}^{phy}\}$ space, we need to act the permutation element $(12)$ on $\{\mathcal{O}^{phy}\}$ and then decompose the generated bases back into the combinations of $\{\mathcal{O}^{phy}\}$ bases,
\eq{ \label{eq:Dd_lower}
    (12)(\mathcal{O}^{phy})^D_d=\sum m^{(d-d')}(\mathcal{O}^{phy})^{D'}_{d'},
}
where the superscript  $D$ and $D^\prime$ are fake dimension.
And due to the following reasons,
\begin{itemize}
    \item{}No matter using Schouten Identity or Momentum conservation, it will only lower its fake dim-$D$;
    \item{}$\{\mathcal{O}^{phy}_d\}$ is the complete set of bases with the lowest dimension, so $(12)(\mathcal{O}^{phy}_d)$ could only be decomposed into some lower dimension bases than $d$,
\end{itemize}
each term on RHS of Eq.~(\ref{eq:Dd_lower}) should satisfy
\bea \label{eq:dimdown}
d'\leq d\,, \quad\quad D'\leq D.
\eea
It means the representation matrix $M^{\mathcal{D}}_{(12)}$ is a partitioned upper triangular matrix in the space of the bases arranged in the ascending order of both dim-$D$ and dim-$d$, as shown in Fig.~\ref{fig:Dd_matrix}. This property tells us that the eigenvectors with non-zero eigenvalues of the diagonal sub-matrix in the base space with a certain fake dimension $D$ one to one correspond to the eigenvectors with non-zero eigenvalues of the full matrix $M^{\mathcal{D}}_{(12)}$. So we only need to get these diagonal sub-matrices $M^{\mathcal{D}}_{D(12)}$ in $M^{\mathcal{D}}_{(12)}$ (correspond to the diagonal boxes in Fig.~\ref{fig:Dd_matrix}).
Using the following steps, we can get the sub-matrix  $M^{\mathcal{D}}_{D(12)}$ for a certain dim-$D$ (the box within blue circle in Fig.~\ref{fig:Dd_matrix}).
\begin{itemize}
    \item[1.]Decompose $(12)\{\mathcal{O}^{phy}\}^D$ to $\{\mathcal{B}\cdot H\}^D$, and get the matrix $M^{BH\leftarrow  \mathcal{O}}_{(12)}$
    \item[2.]Decompose $\{\mathcal{O}^{phy}\}^D$ to $\{\mathcal{B}\cdot H\}^D$, and get the matrix $M^{BH\leftarrow \mathcal{O}}_{e}$
    \item[3.]$M^{\mathcal{D}}_{D(12)}=[M^{BH\leftarrow \mathcal{O}}_{e}]^{-1}\times M^{BH\leftarrow \mathcal{O}}_{(12)}$
\end{itemize}
Since $M^{\mathcal{D}}_{D(12)}$ is still partitioned upper triangular matrix,  now we only need to calculate the eigenvectors of its diagonal blocks, which has a certain dim-$D$ and dim-$d$.  When we map an amplitude nominal to an EFT operator, this operator will automatically satisfy identical particle statistics, enforced by Feynman rules. So we do non need to map the full eigenvectors  with non-zero eigenvalues (correspond long polynomials of spinor products) into operator bases but just need to find one independent component of each eigenvector without non-zero and map them into EFT operator, which is the simplest operator bases satisfying spin statistics.

\section{Code Output: 4-pt Gauge Boson EFT Operators}
\label{sec:operator}
In this section, we list all the four-point amplitude bases and the corresponding operator bases for massive gauge boson $Z$ and $W^{\pm}$ at dim-$4$ and $6$. Notice that for simplicity, we do not symmetrize the expressions of the amplitude bases involving identical particles and full expressions can be obtained by just symmetrizing these amplitude bases.   

For the vertex $D^{2n}Z Z W^+ W^-$ ($D_\mu$ is the QED covariant derivative), we have $2$ physical dim-$4$ operators and $29$ dim-$6$ operators, and we define
\eq{
\acute{F}_{\mu\nu}\equiv F_{\mu\nu}+i\tilde{F}_{\mu\nu}
\quad\quad
\grave{F}_{\mu\nu}\equiv F_{\mu\nu}-i\tilde{F}_{\mu\nu}
}
\begin{table}[H]
\begin{tabular}{|l|l|}
\hline
Amplitude & Operator $D^{2n} ZZW^+W^-$ \\ \hline
$\langle\mathbf{13}\rangle\langle\mathbf{24}\rangle[\mathbf{42}][\mathbf{31}]$
&$Z^\mu Z^\nu W^+_\mu W^-_\nu$
\\ \hline

$\langle\mathbf{13}\rangle\langle\mathbf{24}\rangle[\mathbf{43}][\mathbf{21}]$
&$Z^\mu Z_\mu W^{+\nu} W^-_\nu$
\\ \hline
\hline

$\langle\mathbf{34}\rangle[\mathbf{42}][\mathbf{31}][\mathbf{21}]$
&$\acute{Z}^{\mu\nu} Z_{\mu\nu} W^{+\rho} W^-_\rho
    -i \epsilon^{\mu\nu\rho\sigma} Z_{\mu\gamma}Z^\gamma_\nu W^+_\rho W^-_\sigma$
\\ \hline

$\langle\mathbf{24}\rangle[\mathbf{42}][\mathbf{31}]^2$
&$\acute{Z}^{\mu\nu}W^{+}_{\mu\nu}Z^\rho W^-_\rho$
\\ \hline

$\langle\mathbf{24}\rangle[\mathbf{43}][\mathbf{31}][\mathbf{21}]$
&$\acute{Z}^{\mu\nu} (W^+_{\mu\nu} W^{-}_\rho Z^\rho
    +2Z_\mu W^{+}_{\rho\nu}W^{-\rho}
    -2W^-_\mu W^{+}_{\rho\nu}Z^{\rho})$
\\ \hline

$\langle\mathbf{23}\rangle[\mathbf{42}][\mathbf{41}][\mathbf{31}]$
&$\acute{Z}^{\mu\nu} (W^-_{\mu\nu} Z^{\rho} W^+_\rho
    +2W^+_\mu W^{-}_{\rho\nu}Z^{\rho}
    -2Z_\mu W^{-}_{\rho\nu}W^{+\rho})$
\\ \hline

$\langle\mathbf{23}\rangle[\mathbf{43}][\mathbf{41}][\mathbf{21}]$
&$\acute{Z}^{\mu\nu} (W^-_{\mu\nu} W^{+}_\rho Z^\rho
    +2Z_\mu W^{-}_{\rho\nu}W^{+\rho}
    -2W^+_\mu W^{-}_{\rho\nu}Z^{\rho})$
\\ \hline

$\langle\mathbf{12}\rangle[\mathbf{43}][\mathbf{42}][\mathbf{31}]$
&$\acute{W}^{+\mu\nu} W^-_{\mu\nu} Z^{\rho} Z_\rho$
\\ \hline

$\langle\mathbf{24}\rangle\langle\mathbf{34}\rangle[\mathbf{31}][\mathbf{21}]$
&$\acute{Z}^{\mu\nu}( W^-_{\mu\nu} Z^{\rho} W^+_\rho
    -2W^+_\mu W^{-}_{\rho\nu}Z^{\rho}
    -2Z_\mu W^{-}_{\rho\nu}W^{+\rho})$
\\ \hline

$\langle\mathbf{23}\rangle\langle\mathbf{34}\rangle[\mathbf{41}][\mathbf{21}]$
&$\acute{Z}^{\mu\nu}( W^+_{\mu\nu} Z^{\rho} W^-_\rho
    -2W^-_\mu W^+_{\rho\nu}Z^{\rho}
    -2Z_\mu W^+_{\rho\nu}W^{-\rho})$
\\ \hline

$\langle\mathbf{23}\rangle\langle\mathbf{24}\rangle[\mathbf{41}][\mathbf{31}]$
&$\acute{Z}^{\mu\nu}( Z_{\mu\nu} W^{+\rho} W^-_\rho
    -2W^-_\mu Z_{\rho\nu}W^{+\rho}
    -2W^+_\mu Z_{\rho\nu}W^{-\rho})$
\\ \hline

$\langle\mathbf{14}\rangle\langle\mathbf{24}\rangle[\mathbf{32}][\mathbf{31}]$
&$\acute{W}^{+\mu\nu} (W^-_{\mu\nu} Z^{\rho} Z_\rho
    -4Z_\mu W^-_{\rho\nu}Z^{\rho})$
\\ \hline

$\langle\mathbf{13}\rangle\langle\mathbf{23}\rangle[\mathbf{42}][\mathbf{41}]$
&$\acute{W}^{-\mu\nu} ( W^+_{\mu\nu} Z^{\rho} Z_\rho
    -4Z_\mu W^+_{\rho\nu}Z^{\rho})$
\\ \hline

$\langle\mathbf{12}\rangle\langle\mathbf{24}\rangle[\mathbf{43}][\mathbf{31}]$
&$\acute{W}^{+\mu\nu} (  Z_{\mu\nu} W^{-\rho} Z_\rho
    -2Z_\mu Z_{\rho\nu}W^{-\rho}-2W^-_\mu Z_{\rho\nu}Z^{\rho})$
\\ \hline

$\langle\mathbf{12}\rangle\langle\mathbf{23}\rangle[\mathbf{43}][\mathbf{41}]$
&$\acute{W}^{-\mu\nu} ( Z_{\mu\nu} W^{+\rho} Z_\rho
    -2Z_\mu Z_{\rho\nu}W^{+\rho}-2W^+_\mu Z_{\rho\nu}Z^{\rho})$
\\ \hline

$\langle\mathbf{13}\rangle\langle\mathbf{24}\rangle\langle\mathbf{34}\rangle
[\mathbf{21}]$
&$\grave{W}^{+\mu\nu}W^-_{\mu\nu} Z^{\rho} Z_\rho$
\\ \hline

$\langle\mathbf{13}\rangle\langle\mathbf{24}\rangle^2
[\mathbf{31}]$
&$\grave{Z}^{\mu\nu}W^-_{\mu\nu} W^{+\rho} Z_\rho$
\\ \hline

$\langle\mathbf{12}\rangle\langle\mathbf{24}\rangle\langle\mathbf{34}\rangle
[\mathbf{31}]$
&$\grave{Z}^{\mu\nu}(W^-_{\mu\nu} W^{+\rho} Z_\rho
    +2Z_\mu W^{-}_{\rho\nu}W^{+\rho}
    -2W^+_\mu W^{-}_{\rho\nu}Z^{\rho})$
\\ \hline

$\langle\mathbf{13}\rangle\langle\mathbf{23}\rangle\langle\mathbf{24}\rangle
[\mathbf{41}]$
&$\grave{W}^{+\mu\nu}( Z_{\mu\nu} W^{-\rho} Z_\rho
    +2Z_\mu Z_{\rho\nu}W^{-\rho}
    -2W^+_\mu Z_{\rho\nu}Z^{\rho})$
\\ \hline

$\langle\mathbf{12}\rangle\langle\mathbf{23}\rangle\langle\mathbf{34}\rangle
[\mathbf{41}]$
&$\grave{Z}^{\mu\nu}( W^+_{\mu\nu} W^{-\rho} Z_\rho
    +2Z_\mu W^{+}_{\rho\nu}W^{-\rho}
    -2W^-_\mu W^{+}_{\rho\nu}Z^{\rho})$
\\ \hline

$\langle\mathbf{12}\rangle\langle\mathbf{13}\rangle\langle\mathbf{24}\rangle
[\mathbf{43}]$
&$\grave{Z}^{\mu\nu}(Z_{\mu\nu} W^{-\rho} W^+_\rho
    +2W^+_\mu Z_{\rho\nu}W^{-\rho}
    -2W^-_\mu Z_{\rho\nu}W^{+\rho})$
\\ \hline

$\langle\mathbf{3}2\mathbf{3}]\langle\mathbf{24}\rangle
[\mathbf{41}][\mathbf{21}]$
&$i\acute{Z}^{\mu\nu} W^-_\mu (D^\rho Z_\nu)W^+_\rho$
\\ \hline

$\langle\mathbf{4}2\mathbf{4}]\langle\mathbf{23}\rangle
[\mathbf{31}][\mathbf{21}]$
&$i\acute{Z}^{\mu\nu}W^+_\mu (D^\rho Z_\nu) W^-_\rho$
\\ \hline

$\langle\mathbf{1}3\mathbf{4}]\langle\mathbf{24}\rangle
[\mathbf{32}][\mathbf{31}]$
&$iD_\rho(\acute{W}^{+\mu\rho}g^{\nu\sigma}+\acute{W}^{+\rho\sigma}g^{\mu\nu}
    +\acute{W}^{+\rho\nu}g^{\mu\sigma}+\acute{W}^{+\sigma\mu}g^{\nu\rho}
    +\acute{W}^{+\nu\mu}g^{\rho\sigma})W^-_\mu Z_\nu Z_\sigma$
\\ \hline

$\langle\mathbf{2}3\mathbf{4}]\langle\mathbf{13}\rangle
[\mathbf{42}][\mathbf{31}]$
&$i\acute{W}^{-\mu\nu}Z_\nu (D_\mu W^+_\rho)Z^\rho$
\\ \hline

$\langle\mathbf{4}2\mathbf{3}]\langle\mathbf{13}\rangle\langle\mathbf{24}\rangle
[\mathbf{21}]$
&$i(\grave{W}^{-\mu\rho}g^{\nu\sigma}+\grave{W}^{-\rho\sigma}g^{\mu\nu}
    +\grave{W}^{-\rho\nu}g^{\mu\sigma}+\grave{W}^{-\sigma\mu}g^{\nu\rho}
    +\grave{W}^{-\nu\mu}g^{\rho\sigma}+\grave{W}^{-\sigma\nu}g^{\mu\rho})(D_\sigma Z_\nu)Z_\rho W^+_\mu$
\\ \hline

$\langle\mathbf{4}2\mathbf{4}]\langle\mathbf{13}\rangle\langle\mathbf{23}\rangle
[\mathbf{21}]$
&$i\grave{W}^{+\mu\nu}(D_\rho Z_\mu)Z_\nu W^{-\rho}$
\\ \hline

$\langle\mathbf{2}3\mathbf{4}]\langle\mathbf{13}\rangle\langle\mathbf{24}\rangle
[\mathbf{31}]$
&$i\grave{Z}^{\mu\nu}W^-_\mu (D_\nu W^+_\rho)Z^\rho$
\\ \hline

$\langle\mathbf{4}3\mathbf{4}]\langle\mathbf{12}\rangle\langle\mathbf{23}\rangle
[\mathbf{31}]$
&$i\grave{Z}^{\mu\nu}Z_\mu (D_\rho W^+_\nu)W^{-\rho}$
\\ \hline

$\langle\mathbf{2}3\mathbf{2}]\langle\mathbf{4}2\mathbf{3}]\langle\mathbf{13}\rangle
[\mathbf{41}]$
&$(i\epsilon^{\mu\nu\rho\sigma}+g^{\mu\sigma}g^{\nu\rho}-g^{\mu\rho}g^{\nu\sigma}
    +g^{\mu\nu}g^{\rho\sigma})(D_\xi W^+_\mu) W^-_\rho (D_\nu  Z_\xi) Z^\sigma$
\\ \hline

$\langle\mathbf{4}2\mathbf{4}]\langle\mathbf{2}4\mathbf{2}]\langle\mathbf{13}\rangle
[\mathbf{31}]$
&$Z^\rho W^+_\rho (D_\mu W^{-\nu})(D_\nu Z^\mu)$
\\ \hline
\end{tabular}
\end{table}

And for vertex $D^{2n}W^+W^+W^-W^-$, we have $2$ physical dim-$4$ operators, and $18$ dim-$6$ operators,
\begin{table}[H]
\begin{tabular}{|l|l|}
\hline
Amplitude & Operator $D^{2n}W^+W^+W^-W^-$  \\ \hline
$\langle\mathbf{13}\rangle\langle\mathbf{24}\rangle[\mathbf{42}][\mathbf{31}]$
&$W^{+\mu} W^{+\nu} W^-_\mu W^-_\nu$
\\ \hline

$\langle\mathbf{13}\rangle\langle\mathbf{24}\rangle[\mathbf{43}][\mathbf{21}]$
&$W^{+\mu} W^{+}_\mu W^{-\nu} W^-_\nu$
\\ \hline
\hline

$\langle\mathbf{34}\rangle[\mathbf{42}][\mathbf{31}][\mathbf{21}]$
&$\acute{W}^{+\mu\nu} W^+_{\mu\nu} W^{-\rho} W^-_\rho$
\\ \hline

$\langle\mathbf{24}\rangle[\mathbf{42}][\mathbf{31}]^2$
&$\acute{W}^{+\mu\nu}W^{-}_{\mu\nu}W^{+\rho} W^-_\rho$
\\ \hline

$\langle\mathbf{24}\rangle[\mathbf{43}][\mathbf{31}][\mathbf{21}]$
&$\acute{W}^{+\mu\nu} ( W^-_{\mu\nu} W^{-\rho} W^+_\rho
    +2W^+_\mu W^{-}_{\rho\nu}W^{-\rho}
    -2W^-_\mu W^{-}_{\rho\nu}W^{+\rho}$
\\ \hline

$\langle\mathbf{12}\rangle[\mathbf{43}][\mathbf{42}][\mathbf{31}]$
&$\acute{W}^{-\mu\nu} W^-_{\mu\nu} W^{+\rho} W^+_\rho$
\\ \hline

$\langle\mathbf{24}\rangle\langle\mathbf{34}\rangle[\mathbf{31}][\mathbf{21}]$
&$\acute{W}^{+\mu\nu}(W^-_{\mu\nu} W^{+\rho} W^-_\rho
    -2W^-_\mu W^{-}_{\rho\nu}W^{+\rho}
    -2W^+_\mu W^{-}_{\rho\nu}W^{-\rho})$
\\ \hline

$\langle\mathbf{23}\rangle\langle\mathbf{24}\rangle[\mathbf{41}][\mathbf{31}]$
&$\acute{W}^{+\mu\nu}( W^+_{\mu\nu} W^{-\rho} W^-_\rho
    -4W^-_\mu W^{+}_{\rho\nu}W^{-\rho})$
\\ \hline

$\langle\mathbf{14}\rangle\langle\mathbf{24}\rangle[\mathbf{32}][\mathbf{31}]$
&$\acute{W}^{-\mu\nu}( W^-_{\mu\nu} W^{+\rho} W^+_\rho
    -4W^+_\mu W^{-}_{\rho\nu}W^{+\rho})$
\\ \hline

$\langle\mathbf{12}\rangle\langle\mathbf{24}\rangle[\mathbf{43}][\mathbf{31}]$
&$\acute{W}^{-\mu\nu}( W^+_{\mu\nu} W^{-\rho} W^+_\rho
    -2W^+_\mu W^{+}_{\rho\nu}W^{-\rho}
    -2W^-_\mu W^{+}_{\rho\nu}W^{+\rho})$
\\ \hline

$\langle\mathbf{13}\rangle\langle\mathbf{24}\rangle\langle\mathbf{34}\rangle
[\mathbf{21}]$
&$\grave{W}^{-\mu\nu} W^-_{\mu\nu} W^{+\rho} W^+_\rho$
\\ \hline

$\langle\mathbf{13}\rangle\langle\mathbf{24}\rangle^2
[\mathbf{31}]$
&$\grave{W}^{-\mu\nu} W^+_{\mu\nu} W^{-\rho} W^+_\rho$
\\ \hline

$\langle\mathbf{12}\rangle\langle\mathbf{24}\rangle\langle\mathbf{34}\rangle
[\mathbf{31}]$
&$\grave{W}^{+\mu\nu}( W^-_{\mu\nu} W^{-\rho} W^+_\rho
    +2W^+_\mu W^-_{\rho\nu}W^{-\rho}
    -2W^+_\mu W^-_{\rho\nu}W^{+\rho})$
\\ \hline

$\langle\mathbf{12}\rangle\langle\mathbf{13}\rangle\langle\mathbf{24}\rangle
[\mathbf{43}]$
&$\grave{W}^{+\mu\nu}W^+_{\mu\nu} W^{-\rho} W^-_\rho$
\\ \hline

$\langle\mathbf{3}2\mathbf{3}]\langle\mathbf{24}\rangle
[\mathbf{41}][\mathbf{21}]$
&$i\acute{W}^{+\mu\nu} W^-_\mu (D^\rho W^+_\nu)W^-_\rho$
\\ \hline

$\langle\mathbf{1}3\mathbf{4}]\langle\mathbf{24}\rangle
[\mathbf{32}][\mathbf{31}]$
&$iD_\rho(\acute{W}^{-\mu\rho}g^{\nu\sigma}+\acute{W}^{-\rho\sigma}g^{\mu\nu}
    +\acute{W}^{-\rho\nu}g^{\mu\sigma}+\acute{W}^{-\sigma\mu}g^{\nu\rho}
    +\acute{W}^{-\nu\mu}g^{\rho\sigma})W^-_\mu W^+_\nu W^+_\sigma$
\\ \hline

$\langle\mathbf{4}2\mathbf{3}]\langle\mathbf{13}\rangle\langle\mathbf{24}\rangle
[\mathbf{21}]$
&$i(\grave{W}^{-\mu\rho}g^{\nu\sigma}+\grave{W}^{-\rho\sigma}g^{\mu\nu}
    +\grave{W}^{-\rho\nu}g^{\mu\sigma}+\grave{W}^{-\sigma\mu}g^{\nu\rho}
    +\grave{W}^{-\nu\mu}g^{\rho\sigma}+\grave{W}^{-\sigma\nu}g^{\mu\rho})(D_\sigma W^+_\nu)W^+_\rho W^-_\mu$
\\ \hline

$\langle\mathbf{2}3\mathbf{4}]\langle\mathbf{13}\rangle\langle\mathbf{24}\rangle
[\mathbf{31}]$
&$i\grave{W}^{+\mu\nu}W^-_\mu (D_\nu W^-_\rho)W^{+\rho}$
\\ \hline

$\langle\mathbf{2}3\mathbf{2}]\langle\mathbf{4}2\mathbf{3}]\langle\mathbf{13}\rangle
[\mathbf{41}]$
&$(i\epsilon^{\mu\nu\rho\sigma}+g^{\mu\sigma}g^{\nu\rho}-g^{\mu\rho}g^{\nu\sigma}
    +g^{\mu\nu}g^{\rho\sigma})(D_\xi W^-_\mu) W^-_\rho (D_\nu  W^{+\xi}) W^{+}_\sigma$
\\ \hline

$\langle\mathbf{4}2\mathbf{4}]\langle\mathbf{2}4\mathbf{2}]\langle\mathbf{13}\rangle
[\mathbf{31}]$
&$W^{+\rho} W^-_\rho (D_\mu W^{-\nu})(D_\nu W^{+\mu})$
\\ \hline
\end{tabular}
\end{table}

Finally, for vertex $D^{2n}ZZZZ$, we only have $1$ physical dim-$4$ operator, and $4$ dim-$6$ operators,
\begin{table}[H]
\begin{tabular}{|l|l|}
\hline
Amplitude & Operator $D^{2n}ZZZZ$ \\ \hline
$\langle\mathbf{13}\rangle\langle\mathbf{24}\rangle[\mathbf{42}][\mathbf{31}]$
&$Z^\mu Z^\nu Z_\mu Z_\nu$
\\ \hline
\hline

$\langle\mathbf{34}\rangle[\mathbf{42}][\mathbf{31}][\mathbf{21}]$
&$\acute{Z}^{\mu\nu} Z_{\mu\nu} Z^{\rho} Z_\rho$
\\ \hline

$\langle\mathbf{24}\rangle\langle\mathbf{34}\rangle[\mathbf{31}][\mathbf{21}]$
&$\acute{Z}^{\mu\nu}( Z_{\mu\nu} Z^{\rho} Z_\rho
    -4Z_\mu Z_{\rho\nu}Z^{\rho})$
\\ \hline

$\langle\mathbf{13}\rangle\langle\mathbf{24}\rangle\langle\mathbf{34}\rangle
[\mathbf{21}]$
&$\grave{Z}^{\mu\nu} Z_{\mu\nu} Z^{\rho} Z_\rho$
\\ \hline

$\langle\mathbf{2}3\mathbf{2}]\langle\mathbf{4}2\mathbf{3}]\langle\mathbf{13}\rangle
[\mathbf{41}]$
&$(D_\mu Z^\nu)  (D_\nu  Z^{\mu}) Z_\rho Z^\rho$
\\ \hline

\end{tabular}
\end{table}


\begin{thebibliography}{99}
\bibitem{Weinberg:1966kf}
S.~Weinberg,
Phys. Rev. Lett. \textbf{17}, 616-621 (1966)
doi:10.1103/PhysRevLett.17.616

\bibitem{Weinberg:1968de}
S.~Weinberg,
Phys. Rev. \textbf{166}, 1568-1577 (1968)
doi:10.1103/PhysRev.166.1568

\bibitem{Weinberg:1978kz}
S.~Weinberg,
Physica A \textbf{96}, no.1-2, 327-340 (1979)
doi:10.1016/0378-4371(79)90223-1



\bibitem{Gasser:1982ap}
J.~Gasser and H.~Leutwyler,
Phys. Rept. \textbf{87}, 77-169 (1982)
doi:10.1016/0370-1573(82)90035-7

\bibitem{Gasser:1983yg}
J.~Gasser and H.~Leutwyler,
Annals Phys. \textbf{158}, 142 (1984)
doi:10.1016/0003-4916(84)90242-2

\bibitem{Falkowski:2019tft}
A.~Falkowski and R.~Rattazzi,
JHEP \textbf{10} (2019), 255
doi:10.1007/JHEP10(2019)255
[arXiv:1902.05936 [hep-ph]].

\bibitem{Cohen:2020xca}
T.~Cohen, N.~Craig, X.~Lu and D.~Sutherland,
JHEP \textbf{03} (2021), 237
doi:10.1007/JHEP03(2021)237
[arXiv:2008.08597 [hep-ph]].

\bibitem{Goodman:2010ku}
J.~Goodman, M.~Ibe, A.~Rajaraman, W.~Shepherd, T.~M.~P.~Tait and H.~B.~Yu,
Phys. Rev. D \textbf{82}, 116010 (2010)
doi:10.1103/PhysRevD.82.116010
[arXiv:1008.1783 [hep-ph]].

\bibitem{Cao:2009uw}
Q.~H.~Cao, C.~R.~Chen, C.~S.~Li and H.~Zhang,
JHEP \textbf{08}, 018 (2011)
doi:10.1007/JHEP08(2011)018
[arXiv:0912.4511 [hep-ph]].

\bibitem{Zheng:2010js}
J.~M.~Zheng, Z.~H.~Yu, J.~W.~Shao, X.~J.~Bi, Z.~Li and H.~H.~Zhang,
Nucl. Phys. B \textbf{854} (2012), 350-374
doi:10.1016/j.nuclphysb.2011.09.009
[arXiv:1012.2022 [hep-ph]].

\bibitem{Aebischer:2022wnl}
J.~Aebischer, W.~Altmannshofer, E.~E.~Jenkins and A.~V.~Manohar,
[arXiv:2202.06968 [hep-ph]].


\bibitem{Jenkins:2017jig}
E.~E.~Jenkins, A.~V.~Manohar and P.~Stoffer,
JHEP \textbf{03}, 016 (2018)
doi:10.1007/JHEP03(2018)016
[arXiv:1709.04486 [hep-ph]].


\bibitem{Bern:2020ikv}
Z.~Bern, J.~Parra-Martinez and E.~Sawyer,
JHEP \textbf{10}, 211 (2020)
doi:10.1007/JHEP10(2020)211
[arXiv:2005.12917 [hep-ph]].

\bibitem{Jiang:2020mhe}
M.~Jiang, T.~Ma and J.~Shu,
JHEP \textbf{01}, 101 (2021)
doi:10.1007/JHEP01(2021)101
[arXiv:2005.10261 [hep-ph]].


\bibitem{EliasMiro:2020tdv}
J.~Elias Mir\'o, J.~Ingoldby and M.~Riembau,
JHEP \textbf{09}, 163 (2020)
doi:10.1007/JHEP09(2020)163
[arXiv:2005.06983 [hep-ph]].


\bibitem{Baratella:2020lzz}
P.~Baratella, C.~Fernandez and A.~Pomarol,
Nucl. Phys. B \textbf{959}, 115155 (2020)
doi:10.1016/j.nuclphysb.2020.115155
[arXiv:2005.07129 [hep-ph]].

\bibitem{Baratella:2020dvw}
P.~Baratella, C.~Fernandez, B.~von Harling and A.~Pomarol,
JHEP \textbf{03}, 287 (2021)
doi:10.1007/JHEP03(2021)287
[arXiv:2010.13809 [hep-ph]].


\bibitem{Shu:2021qlr}
J.~Shu, M.~L.~Xiao and Y.~H.~Zheng,
[arXiv:2111.08019 [hep-th]].


\bibitem{AccettulliHuber:2021uoa}
M.~Accettulli Huber and S.~De Angelis,
JHEP \textbf{11}, 221 (2021)
doi:10.1007/JHEP11(2021)221
[arXiv:2108.03669 [hep-th]].





\bibitem{Cheung:2015aba}
  C.~Cheung and C.~H.~Shen,
  Phys.\ Rev.\ Lett.\  {\bf 115}, no. 7, 071601 (2015)
  doi:10.1103/PhysRevLett.115.071601
  [arXiv:1505.01844 [hep-ph]].


\bibitem{Jiang:2020rwz}
M.~Jiang, J.~Shu, M.~L.~Xiao and Y.~H.~Zheng,
Phys. Rev. Lett. \textbf{126}, no.1, 011601 (2021)
doi:10.1103/PhysRevLett.126.011601
[arXiv:2001.04481 [hep-ph]].

\bibitem{Rose:2022njd}
L.~D.~Rose, B.~von Harling and A.~Pomarol,
[arXiv:2201.10572 [hep-ph]].


\bibitem{Cheung:2014dqa}
C.~Cheung, K.~Kampf, J.~Novotny and J.~Trnka,
Phys. Rev. Lett. \textbf{114}, no.22, 221602 (2015)
doi:10.1103/PhysRevLett.114.221602
[arXiv:1412.4095 [hep-th]].

\bibitem{Cheung:2016drk}
C.~Cheung, K.~Kampf, J.~Novotny, C.~H.~Shen and J.~Trnka,
JHEP \textbf{02}, 020 (2017)
doi:10.1007/JHEP02(2017)020
[arXiv:1611.03137 [hep-th]].

\bibitem{Low:2014nga}
I.~Low,
Phys. Rev. D \textbf{91}, no.10, 105017 (2015)
doi:10.1103/PhysRevD.91.105017
[arXiv:1412.2145 [hep-th]].

\bibitem{Low:2014oga}
I.~Low,
Phys. Rev. D \textbf{91}, no.11, 116005 (2015)
doi:10.1103/PhysRevD.91.116005
[arXiv:1412.2146 [hep-ph]].


\bibitem{Elvang:2010jv}
  H.~Elvang, D.~Z.~Freedman and M.~Kiermaier,
  JHEP {\bf 1011}, 016 (2010)
  doi:10.1007/JHEP11(2010)016
  [arXiv:1003.5018 [hep-th]].

\bibitem{Shadmi:2018xan}
  Y.~Shadmi and Y.~Weiss,
  arXiv:1809.09644 [hep-ph].

\bibitem{Ma:2019gtx}
T.~Ma, J.~Shu and M.~L.~Xiao,
[arXiv:1902.06752 [hep-ph]].

\bibitem{Falkowski:2019zdo}
A.~Falkowski,
[arXiv:1912.07865 [hep-ph]].







\bibitem{Henning:2019enq}
B.~Henning and T.~Melia,
Phys. Rev. D \textbf{100}, no.1, 016015 (2019)
doi:10.1103/PhysRevD.100.016015
[arXiv:1902.06754 [hep-ph]].



\bibitem{Li:2020gnx}
H.~L.~Li, Z.~Ren, J.~Shu, M.~L.~Xiao, J.~H.~Yu and Y.~H.~Zheng,
Phys. Rev. D \textbf{104}, no.1, 015026 (2021)
doi:10.1103/PhysRevD.104.015026
[arXiv:2005.00008 [hep-ph]].

\bibitem{Li:2020zfq}
H.~L.~Li, J.~Shu, M.~L.~Xiao and J.~H.~Yu,
[arXiv:2012.11615 [hep-ph]].


\bibitem{Durieux:2019eor}
G.~Durieux, T.~Kitahara, Y.~Shadmi and Y.~Weiss,
JHEP \textbf{01}, 119 (2020)
doi:10.1007/JHEP01(2020)119
[arXiv:1909.10551 [hep-ph]].


\bibitem{Durieux:2020gip}
G.~Durieux, T.~Kitahara, C.~S.~Machado, Y.~Shadmi and Y.~Weiss,
JHEP \textbf{12}, 175 (2020)
doi:10.1007/JHEP12(2020)175
[arXiv:2008.09652 [hep-ph]].

\bibitem{Falkowski:2020fsu}
A.~Falkowski, G.~Isabella and C.~S.~Machado,
SciPost Phys. \textbf{10}, no.5, 101 (2021)
doi:10.21468/SciPostPhys.10.5.101
[arXiv:2011.05339 [hep-ph]].

\bibitem{Balkin:2021dko}
R.~Balkin, G.~Durieux, T.~Kitahara, Y.~Shadmi and Y.~Weiss,
[arXiv:2112.09688 [hep-ph]].


\bibitem{Dong:2021yak}
Z.~Y.~Dong, T.~Ma and J.~Shu,
[arXiv:2103.15837 [hep-ph]].


\bibitem{Arkani-Hamed:2017jhn}
  N.~Arkani-Hamed, T.~C.~Huang and Y.~t.~Huang,
  arXiv:1709.04891 [hep-th].











\bibitem{Donziyu}
Z.~Y.~Dong, T.~Ma, J.~Shu, Z.~Z.~Zhou, to appear.

\bibitem{DeAngelis:2022qco}
S.~De Angelis,
[arXiv:2202.02681 [hep-th]].


\end{thebibliography}
\end{document}